\documentclass[pra,aps,twocolumn,superscriptaddress,showpacs]{revtex4}

\usepackage{epsfig}
\usepackage{amsmath}
\usepackage[mathscr]{eucal}

\newcommand{\e}{\ensuremath{\mathrm{e}}}
\newcommand{\im}{\ensuremath{\mathrm{i}}}
\newcommand{\tm}{\ensuremath{t_{\mathrm{m}}}}
\newcommand{\VSD}{\ensuremath{V_{\mathrm{SD}}}}

\usepackage{grffile}

\usepackage{ulem}
\usepackage{color}
\definecolor{darkblue}{rgb}{0.0,0,0.6}
\definecolor{darkred}{rgb}{0.6,0,0.0}

\begin{document}

\title{Full counting statistics in the not-so-long-time limit}

\author{Sam T. Carr}
\affiliation{School of Physical Sciences, University of Kent, Canterbury CT2 7NH,
UK}

\author{Peter Schmitteckert}
\affiliation{DFG Center for Functional Nanostructures, Karlsruhe Institute of Technology, 76128 Karlsruhe, Germany}
\affiliation{Institute of Nanotechnology, Karlsruhe Institute of Technology, 76344 Eggenstein-Leopoldshafen, Germany}

\author{Hubert Saleur}
\affiliation{Institut de Physique Th\'eorique, CEA, IPhT and CNRS, URA2306, Gif Sur Yvette, F-91191}
\affiliation{Department of Physics, University of Southern California, Los Angeles, CA 90089-0484}

\date{\today}


\begin{abstract}
The full counting statistics of charge transport is the probability distribution $p_n(t_m)$ that $n$ electrons have flown through the system in measuring time $t_m$.  The cumulant generating function (CGF) of this distribution $F(\chi,t_m)$ has been well studied in the long time limit $t_m\rightarrow \infty$, however there are relatively few results on the finite measuring time corrections to this.  In this work, we study the leading finite time corrections to the CGF of interacting Fermi systems with a single transmission channel at zero temperature but driven out of equilibrium by a bias voltage.  We conjecture that the leading finite time corrections are logarithmic in $t_m$ with a coefficient universally related to the long time limit.  We provide detailed numerical evidence for this with reference to the self-dual interacting resonant level model.  This model further contains a phase transition associated with the fractionalisation of charge at a critical bias voltage.  This transition manifests itself technically as branch points in the CGF.  We provide numerical results of the dependence of the CGF on measuring time for model parameters in the vicinity of this transition, and thus identify features in the time evolution associated with the phase transition itself.

\end{abstract}

\maketitle

\section{Introduction}

The thermodynamic limit in which the system size is taken to be infinite is beloved by theorists as many calculations in many-body physics become dramatically simplified in this limit.  This must be contrasted with experimental work, in which one tends to make measurements on systems of a finite size.  For bulk thermodynamic measurements this discrepancy is typically not important, the reason being that Avogadro's number is so large as to be practically indistinguishable from infinity.  For example, phase transitions are strictly speaking only true transitions in the thermodynamic limit where the free energy exhibits singularities, however their signatures becomes arbitrarily sharp in large but finite systems as to make this mathematical technicality irrelevant.

As nanotechnology improves and nano-devices become smaller, finite-size corrections to the thermodynamic limit become more relevant.  The same is true as computers improve and computational approaches become more commonplace.  Particularly for strongly correlated systems, one often turns to methods based on exact diagonalization or stochastic sampling of the partition function (monte-carlo methods), which tend to be limited to relatively small (certainly compared to Avogardo's number) system sizes.  To extrapolate from these results on small systems to systems of experimental relevance requires an understanding of the finite-size scaling; even typical experimental nano-devices while tiny compared to bulk systems are large compared to the number of particles/sites that may be simulated in the fully quantum methods necessary for understanding strongly correlated systems.

 However one of the more fascinating results of modern theoretical physics is that, at least in a large class of models, the finite-size scaling often reveals more information about the system in question than the bulk result itself.  The most beautiful example of this occurs in systems with conformal invariance \cite{CardyBook,CFT-book}, where the ground state energy of the system per unit length $E_0(L)$ of a system of length $L$ follows the relationship \cite{Bloete-1986,Affleck-1986}
 \begin{equation}
 E_0(L)=E_0(\infty)-\frac{\pi c}{6L}.
 \end{equation}
The crucial thing in this relationship is that the coefficient $c$ appearing in the $1/L$ finite-size scaling term is the central charge of the system.   This is a measure of the number of gapless degrees of freedom in the bulk system, and can not be simply determined from knowledge of $E_0(\infty)$ alone.

When one turns to transport through non-equilibrium nano-structures, one finds a similar dichotomy between theory and experiment in the time domain.  This is most apparent in the study of full counting statistics (FCS) of charge transfer \cite{LL,Levitov-Lee-Lesovik-1996,Nazarov-Blanter-book,Belzig-lecture,Klich-2003,Bagrets-Nazarov-2003}.  The FCS is the study of the full probability distribution $p_n(t_m)$ that $n$ electrons have moved from the source lead to the drain lead in the measuring time $t_m$, typically when driven by a bias voltage $V_\mathrm{SD}$.  This is a probability distribution as the exact number measured in any given realization is subject to both quantum and thermal fluctuations.  In a typical experimental measurement of the FCS \cite{Gustavsson-2006,Gustavsson-2007,Choi-2012}, one counts (having first found a way to do so which is not a trivial matter, but is not the subject of this manuscript) the number of electrons that has flown through the nano-structure in a given time, then repeats the experiment many times to get the distribution.

It is clear that in the long-time limit, the average number $\bar{n}$ of electrons moved, and by extension all moments of the distribution $p_n(t_m)$, should be proportional to the measuring time $t_m$, and indeed the theoretical work has mostly concentrated on the long-time limit $t_m\rightarrow\infty$.  However in the experiments to date \cite{Gustavsson-2006,Gustavsson-2007,Choi-2012}, the source-drain bias voltage $V_{\mathrm{SD}}$ is so low that the relevant dimensionless parameter $eV_\mathrm{SD}t_m/\hbar$ is actually rather small.  In addition, modern numerical techniques to determine the FCS in strongly correlated systems via real-time simulations \cite{CBS-2011} are computationally very expensive and as such can only simulate the system for relatively short measurement times.  One may therefore ask the question: can one understand the contributions to the FCS that are sub-leading in measurement time, and do such contributions give us useful information about the system in question?

Early work in this direction \cite{Muzykantskii-2003,Braunecker-2006,Schoenhammer-2007,Hassler-2008} has concentrated on non-interacting systems; more recently the present authors have presented a conjecture that extends this work to interacting (and in particular strongly correlated) systems \cite{Schmitteckert-Carr-Saleur-2014}, although this result is limited to zero temperature.  In these Proceedings, we review this conjecture along with its background, and present the latest numerical evidence for the conjecture.  In addition, we discuss more general questions about finite-time corrections to FCS, particularly with reference to the measurement of fractionally charged quasi-particles.  To be concrete, we discuss all of this with specific reference to the interacting resonant level model.

The structure of the paper is as follows: in section \ref{sec:FCS-NI}, we review the background of full counting statistics in non-interacting systems in the scattering matrix (Landauer-B\"uttiker) approach.  In section \ref{sec:FCS-I}
we conjecture how this may be extended to interacting systems, giving numerical evidence in support of our conjecture.  In section \ref{sec:FCS-PT} we then discuss one of the more fascinating properties of the interacting resonant level model -- a non-equilibrium phase transition characterised by a bifurcation in the cumulant generating function corresponding to fractionalisation of the charge-carrying quasi-particles.  In particular, we show that the conjecture about finite-time corrections must break down at this critical bias voltage, and the relevance of this to the measurement of fractional charge.  Finally in section \ref{sec:end}, we summarise our results and open questons.

Throughout the paper, we use units where Planck's constant $h=1$, and the unit of electrical charge $e=1$ unless otherwise specified.  Note the slightly unusual choice to use units in which Planck's constant ($h$) rather than the reduced  constant ($\hbar$) is one.  Using this choice, the fundamental unit of conductance $e^2/h$ becomes unity in the dimensionless units, which is convenient for the quantum transport phenomena described in these proceedings.

\section{Full counting statistics in non-interacting systems}
\label{sec:FCS-NI}

\subsection{Basics}

Rather than study directly the probability distribution $p_n(t_m)$ of $n$ electrons (or in other words, a charge of $Q=ne=n$ in our dimensionless units) being moved from the source lead to the drain lead in measuring time $t_m$, it is often more convenient to take the Fourier transform which gives the generating function of the distribution
\begin{equation}
\label{eq:Zdef}
Z(\chi,t_m) = \sum_n e^{i\chi n} p_n = \langle e^{i\chi \hat{Q}} \rangle.
\end{equation}
Here, $\chi$ is known as the counting field, and it is usual to further take the logarithm of the above expression to obtain the \textit{cumulant generating function} (CGF),
\begin{equation}
F(\chi,t_m) =  \ln Z(\chi,t_m).
\label{eq:CGFdef}
\end{equation}
The derivatives of the CGF give the irreducible moments
\begin{equation}
C_n = \left. \left( \frac{\partial}{i\partial\chi} \right)^n F(\chi,t_m) \right|_{\chi=0}.
\end{equation}
Substituting the definition in Eqs.~\eqref{eq:Zdef} and \eqref{eq:CGFdef} into this expression, we see that the first cumulant
\begin{equation}
C_1 = \langle \hat{Q} \rangle \sim I t_m
\end{equation}
where $I$ is the current flowing through the system, while the second cumulant
\begin{equation}
C_2 = \langle \hat{Q}^2 \rangle - \langle \hat{Q} \rangle^2 \sim S t_m
\end{equation}
where $S$ is the zero-frequency shot noise.  In fact to make the above expressions rigorous in quantum mechanics, one must supplement the definition \eqref{eq:Zdef} with a prescription for the appropriate time-ordering of the operator $\hat{Q}=\int_0^{t_m} dt \hat{I}(t)$ as the current operator doesn't commute with itself at different times; we refer to the literature \cite{LL,Levitov-Lee-Lesovik-1996,Belzig-lecture,CBS-2011} for a full discussion on this.

The CGF encodes all information about the statistics of charge transfer through the nanostructure, however one of the easiest quantities to obtain from the CGF is the charge on the (quasi)particles involved in transport.  From the basic definition \eqref{eq:Zdef}, one sees that if only single charges $q=1$ ($q=e$ if units are restored) may be transferred between the leads, then $F(\chi)$ must be a $2\pi$ periodic function.  If however one has, for example, a superconductor where only cooper pairs may be transferred with charge $q=2$, one will see that $F(\chi+\pi)=F(\chi)$ \cite{Belzig-lecture}.  More interestingly, if one sees a periodicity larger than $2\pi$, then this is an indication that the particles involved in transport are fractionally charged \cite{Belzig-lecture,Saleur-Weiss-2001,Levitov-Resnikov-2004}.  Such fractionally charged quasi-particles appear in strongly correlated systems, where the charge may even change with bias voltage \cite{Ivanov-Abanov-2010,CBS-2011}.  We will discuss this point in more detail in section \ref{sec:FCS-PT}.

We add here one final comment about the periodicity of the CGF.  As $F(\chi)$ is given by a complex logarithm, \eqref{eq:CGFdef}, the imaginary part of $F$ is only defined modulo $2\pi$.  The imaginary part of the CGF can therefore contain a linear component $F(\chi) = 2n i \pi \chi$ and still be classed as periodic.  As we will see later, such a linear component is associated with perfect transmission through the system, and thus is often seen in model systems as it has a particular physical significance.

\subsection{Cumulant Generating Function for non interacting systems}

As mentioned earlier, the average charge transferred between the leads in time $t_m$ is expected to grow linearly in measuring time $t_m$, and by extension one expects all cumulants to exhibit similar behavior.  Hence the cumulant generating function is expected in the long time limit $F(\chi,t_m) \sim \tilde{F}_0(\chi) t_m$.  It has been shown that the subleading corrections to this at zero temperature $T=0$ are logarithmic \cite{Muzykantskii-2003,Braunecker-2006,Hassler-2008} in nature, and hence one may write $F(\chi,t_m)$ as a formal expansion in the small parameter $(V_\mathrm{SD} t_m)^{-1}$:
\begin{equation}
F(\chi, \tm ) =\; \tilde{F}_0 t_m + \tilde{F}_1 \ln\left(V_\mathrm{SD}t_m \right) + \cdots
\label{eq:series1}
\end{equation}
For plotting and comparison to numerical work it is more convenient to take the time derivative of this,
\begin{equation}
 \dot F(\chi, \tm ) =  \tilde{F}_0 + \tilde{F}_1/t_m + \cdots \label{eq:series}
\end{equation}
From hereon when we refer to the CGF, we will most typically mean $\dot F(\chi,t_m)$, however this should always be clear from context.

For non-interacting system, it is most convenient to use the scattering (Landauer-B\"uttiker) formalism (see e.g. \cite{Datta_book}) which describes transport properties in terms of the transmission $T(\epsilon)$ through the system.   Within this approach, the  leading term in the expansion of the CGF was derived by Levitov and Lesovik \cite{LL,Levitov-Lee-Lesovik-1996} and is given by
\begin{equation}
\tilde{F_0}(\chi) = \int_{V_D}^{V_S} d\epsilon \ln\left[ 1 + T(\epsilon) \left(e^{i\chi} -1 \right) \right].
\label{eq:LL}
\end{equation}
Here, $V_S$ and $V_D$ are the source and drain voltages respectively, the applied bias $V_\mathrm{SD}=V_S-V_D$.  In order to understand this expression we note that If the transmission is independent of energy $T(\epsilon)\equiv T$, then this gives $\tilde{F}_0 \propto \ln[1+T(e^{i\chi} -1)]$, which is just the CGF for the Bernoulli distribution where each event occurs with probability $T$.  A full introduction to this may be found in Ref.~\onlinecite{Belzig-lecture}.

We now turn our attention to the subleading correction, which has been found \cite{Muzykantskii-2003,Hassler-2008} to be given by
\begin{equation}
\tilde{F}_1(\chi) = \frac{1}{2\pi} \sum_{\epsilon=V_D,V_S} \ln^2 \left[1+T(\epsilon)(\e^{\im\chi} - 1 ) \right].
\label{eq:NI-F1}
\end{equation}
While we refer to the original papers for the technical details of the derivation of this expression \eqref{eq:NI-F1}, it is worth making some comments on the origin of these contributions to the CGF that are logarithmic in measuring time.  The CGF \eqref{eq:CGFdef} may be represented as the determinant of a matrix \cite{Klich} which has a Toeplitz form.  The leading contribution to this determinant is then given by Szego's theorem \cite{Szego}; this is the Levitov-Lesovik result \eqref{eq:LL}.  However it is known that if the matrix elements have certain singularities then the determinant acquires logarithmic corrections according to the Fisher-Hartwig conjecture \cite{FH}.  In this case, the singularities come from the Fermi edges which are infinitely sharp at zero temperature, the correction being Eq.~\eqref{eq:NI-F1} which was first derived in Ref.~\onlinecite{Hassler-2008}.

The crucial point above is that the logarithmic corrections are due the sharp Fermi edges.  The first implication of this is that at non-zero temperature when the Fermi edges are smooth, the long time expansion \eqref{eq:series1} \textit{does not} contain the (non-analytic) logarithmic term.  At low temperatures $T$ however, there is a crossover time $t_X \sim 1/T$; at measurement times $t_m \ll t_X$ the CGF will include the logarithmic terms as if the system were at $T=0$, crossing over at long times $t_m \gg t_X$ to a different temperature-dependent result \cite{Braunecker-2006}.  The second implication is that the coefficient of the logarithmic term, \eqref{eq:NI-F1}, depends on the system \textit{only} in the vicinity of the Fermi edges.  This is clearly true in Eq.~\eqref{eq:NI-F1} where the expression depends on the transmission only at the Fermi energies of the left and right leads (which are different due to the bias voltage).  

We emphasize this point as it will turn out to be very important when we generalize these results to interacting systems in section \ref{sec:FCS-I}.  Before we do this however, we look at an example of the evolution of the CGF as a function of measuring time in a non-interacting system.

\subsection{Application to the non-interacting resonant level model}

\begin{figure}
\begin{center}
\includegraphics[width=\columnwidth]{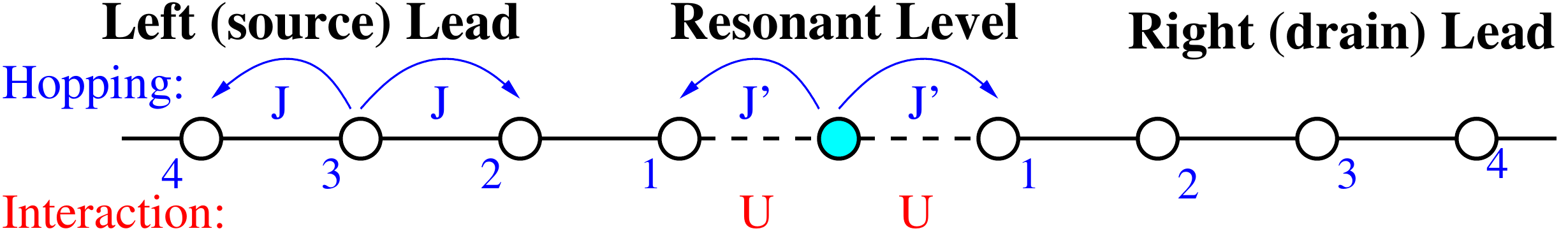}
\end{center}
\caption{A schematic representation of the interacting resonant level model.  If the interaction $U=0$, then the system is the non-interacting resonant level model.}
\label{fig:IRLM_lattice}
\end{figure}

The resonant level model (RLM) is a fermionic model of a single (spinless) state hybridized with two leads, which for convenience are modelled as tight-binding chains.  The Hamiltonian of the system, which is represented schematically in Fig.~\ref{fig:IRLM_lattice} is given by
\begin{multline}
{\cal H}_\mathrm{RLM} = \sum_{a=L,R} \left\{ -J \sum_{n=0}^{M_a} \left( \hat{c}_{a,n}^\dagger \hat{c}_{a,n+1} + H.c. \right)  \right.  \\
\left. -J' \hat{c}_{a,0}^\dagger \hat{d} + H.c. \right\}
\label{eq:RLM}
\end{multline}
In this expression, $\hat{c}_{a,n}^\dagger$ is the fermionic creation operator on lead $a$ (the left and right leads correspond to the source and drain in the transport set up) and site $n$, while $\hat{d}^\dagger$ is the creation operator on the resonant level.  The hopping parameter of the leads is taken to be $J$ (this is conventionally notated as $t$ in condensed matter, however we find $J$ more convenient in non-equilibrium situations when $t$ is the time), and the length of each lead is $M_a$.  While this is typically taken to be infinite in analytic calculations, numerical simulations are done with finite $M_a$, see Appendix \ref{sec:Size}.  Finally, the hybridization between the resonant level and the leads is given by $J'$.

\begin{figure*}
\begin{center}
\includegraphics[width=0.49\textwidth,clip=true]{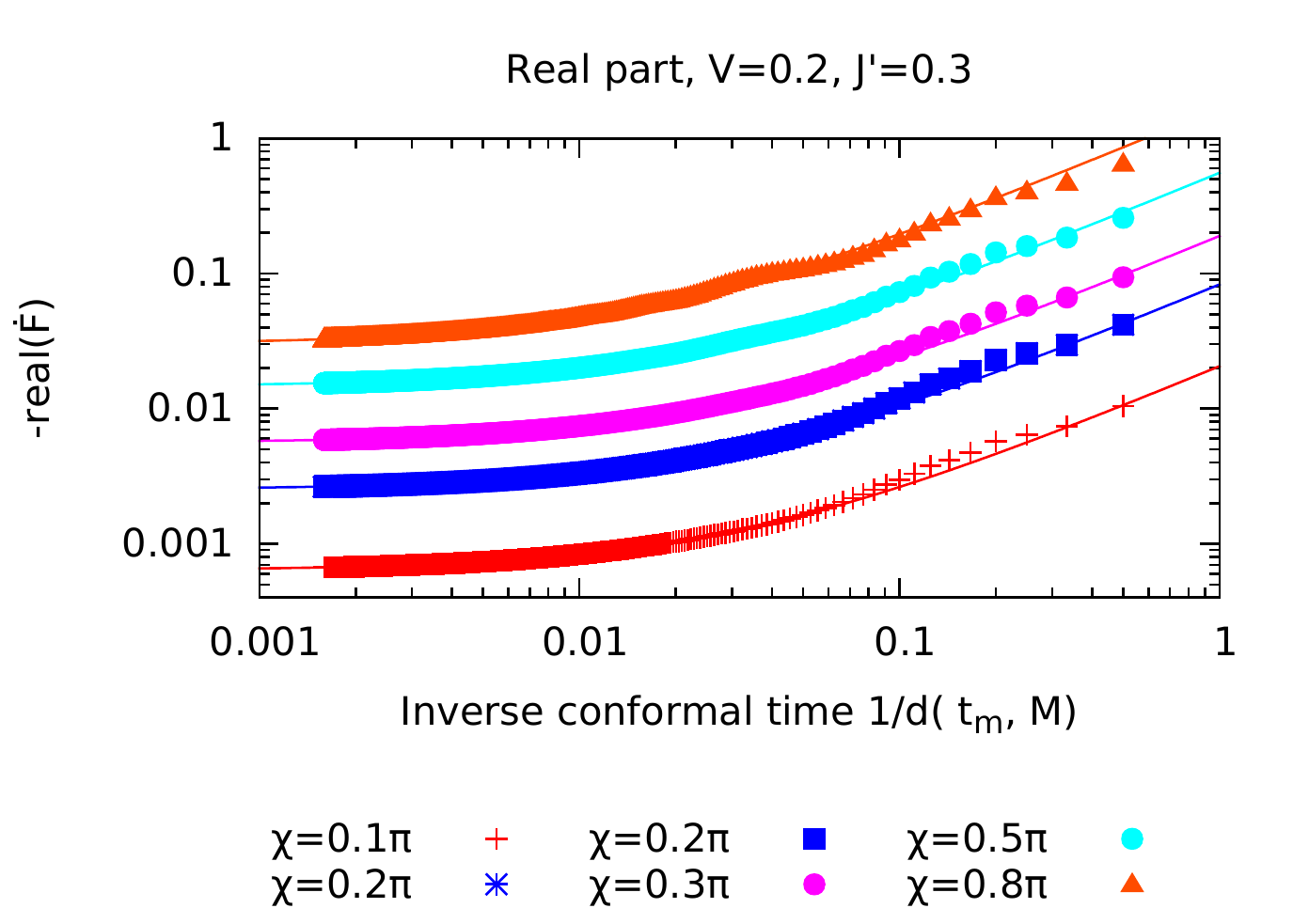}
\includegraphics[width=0.49\textwidth,clip=true]{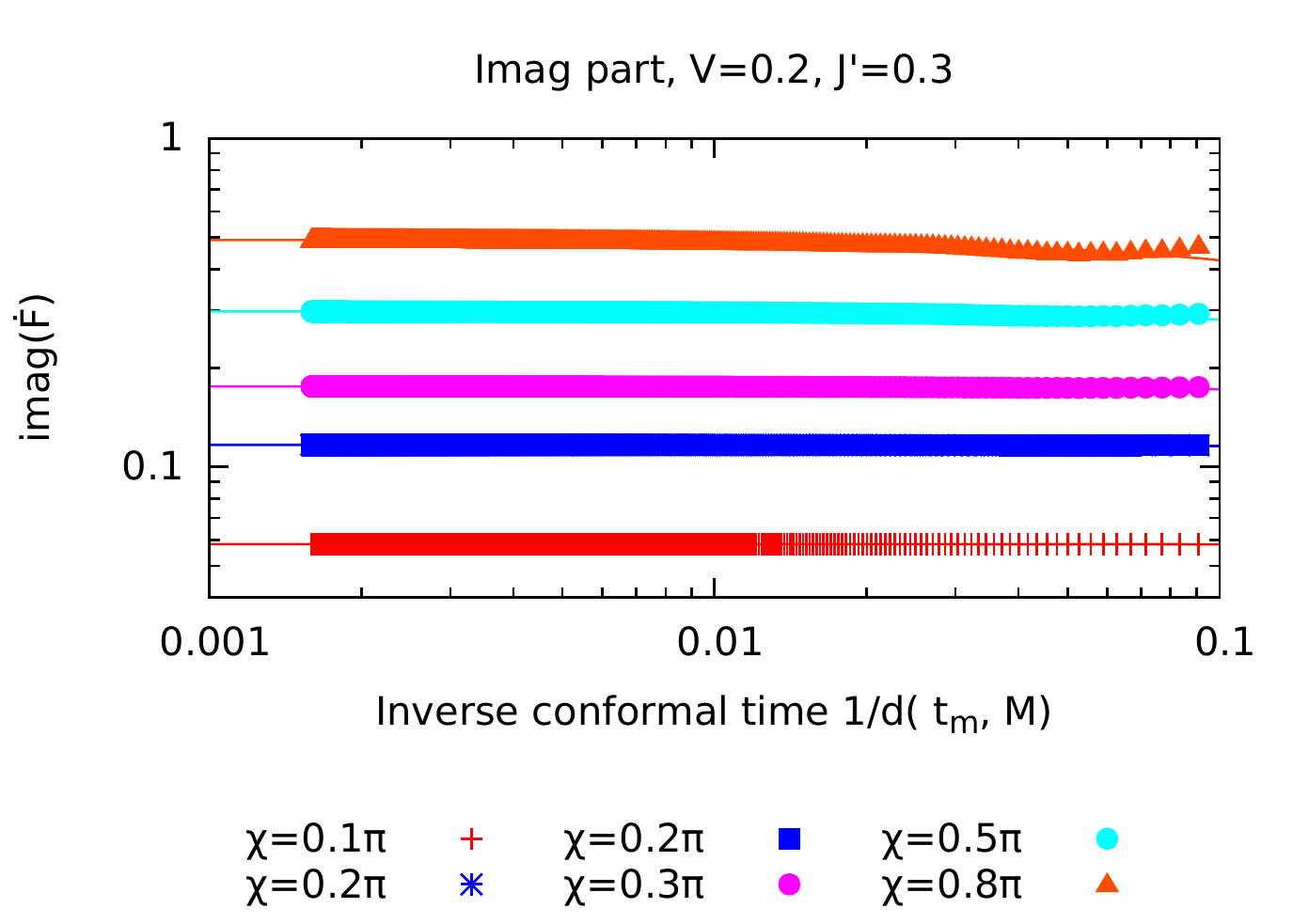}
\end{center}
\caption{Real part (left panel) and imaginary part (right panel) of the CGF for the RLM as a function of measuring time various values of the counting field $\chi$.  
The symbols are the results from real time numerical simulation with system size $M=4000$, while the lines are from the analytic result \eqref{eq:series} with the conformal time substitution (see main text and Appendix \ref{sec:Size} for details).   The coefficients of the series $\tilde{F}_{0,1}$ given by Eqs.~\eqref{eq:F0-RLM} and \eqref{eq:F1-RLM}.
}\label{fig:RLM-time-evolution}
\end{figure*}

As this is a non-interacting problem, the transport through the resonant level when a bias voltage is applied between the two leads may be calculated in the scattering matrix formalism.  The transmission $T(\epsilon)$ through the dot as a function of energy $\epsilon$ in the RLM, \eqref{eq:RLM} is given by (see e.g. \cite{Branschaedel_Boulat_Saleur_Schmitteckert:PRB2010})
\begin{equation}
T(\epsilon) = \frac{1-a^2\epsilon^2}{1+b^2\epsilon^2} \label{eq:T-RLM}
\end{equation}
where $a^2 = 1/4J^2$ and $b^2=\frac{J^2-2J'^2}{4J'^4}$.  This equation is for the specific model \eqref{eq:RLM} which has tight-binding leads with a cosine band.  One often takes the limit of a wide band, corresponding to $\epsilon,J' \ll J$ which gives the more usual Lorentzian shape of the transmission
\begin{equation*}
T(\epsilon) \approx \frac{1}{1+\epsilon^2/\Gamma^2}
\end{equation*}
where $\Gamma=2J'^2/J$.  For comparison with high-precision numerics however, it is necessary to keep the full structure of \eqref{eq:T-RLM}.

Substituting the expression \eqref{eq:T-RLM} into the general Levitov-Lesovik formula, \eqref{eq:LL} and performing the integral gives us the leading (in inverse measuring time) term of the CGF:
\begin{eqnarray}
\tilde{F}_0(\chi) &=& \VSD\ln \left[ 1+ \frac{(e^{\im\chi}-1)(1-a^2\VSD^2/4)}{1+b^2\VSD^2/4}\right] \nonumber \\
&&+ \frac{4e^{\im\chi/2}\tan^{-1}\left[\frac{\VSD}{2}e^{-\im\chi/2}\sqrt{b^2-a^2(e^{\im\chi}-1)}\right]}{\sqrt{b^2-a^2(e^{\im\chi}-1)}} \nonumber \\
&&- \frac{4\tan^{-1}(b\VSD/2)}{b}. \label{eq:F0-RLM}
\end{eqnarray}
The subleading term is then determined from \eqref{eq:NI-F1} to be
\begin{equation}
\tilde{F}_1(\chi)=\frac{1}{\pi}  \ln^2 \left(1+ \frac{1-\VSD^2/16}{1+\frac{1-2J'^2}{16J'^4}\VSD^2}  (\e^{\im\chi} - 1 ) \right).\label{eq:F1-RLM}
\end{equation}

\begin{figure*}
\begin{center}
\includegraphics[width=0.49\textwidth,clip=true]{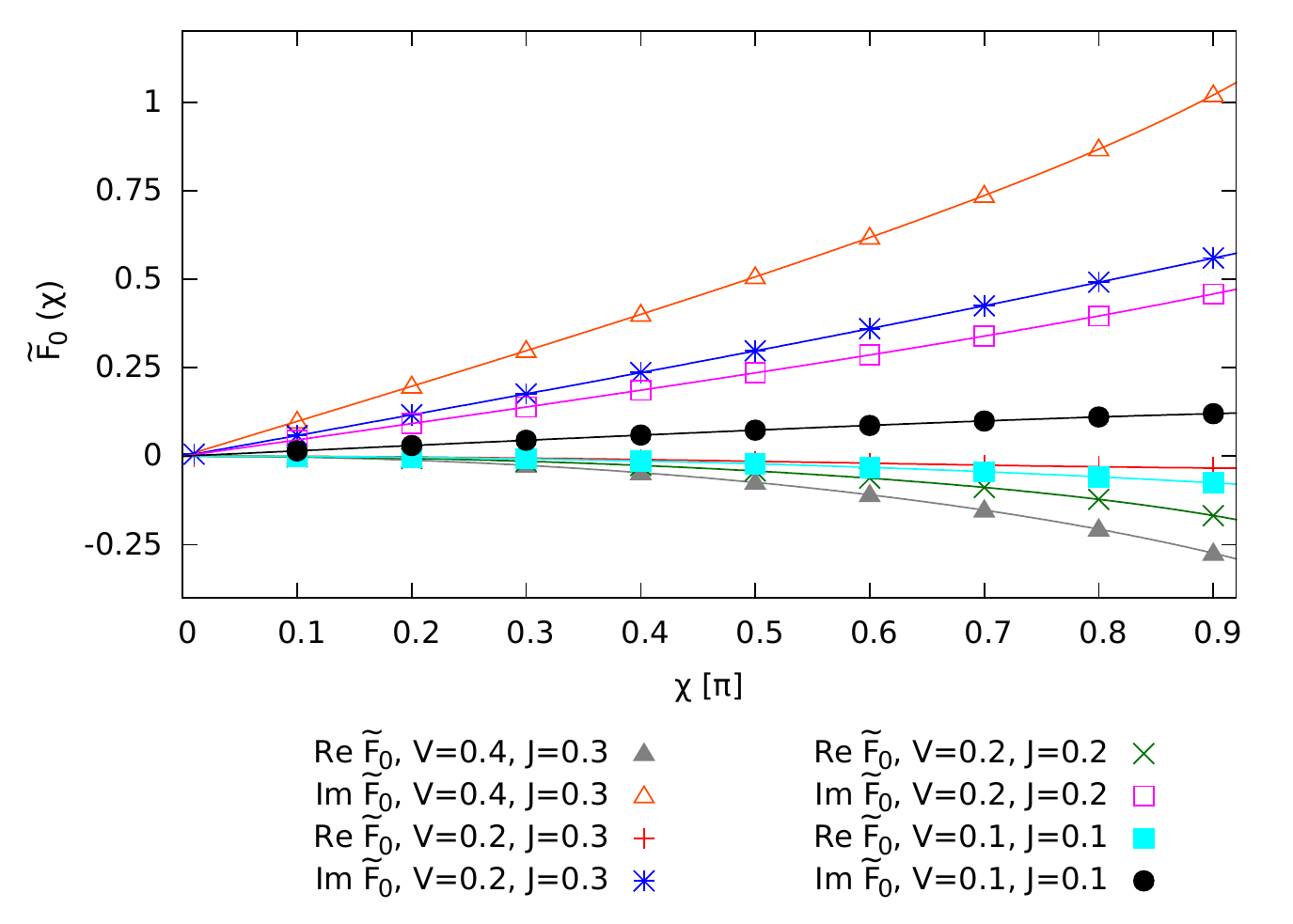}
\includegraphics[width=0.49\textwidth,clip=true]{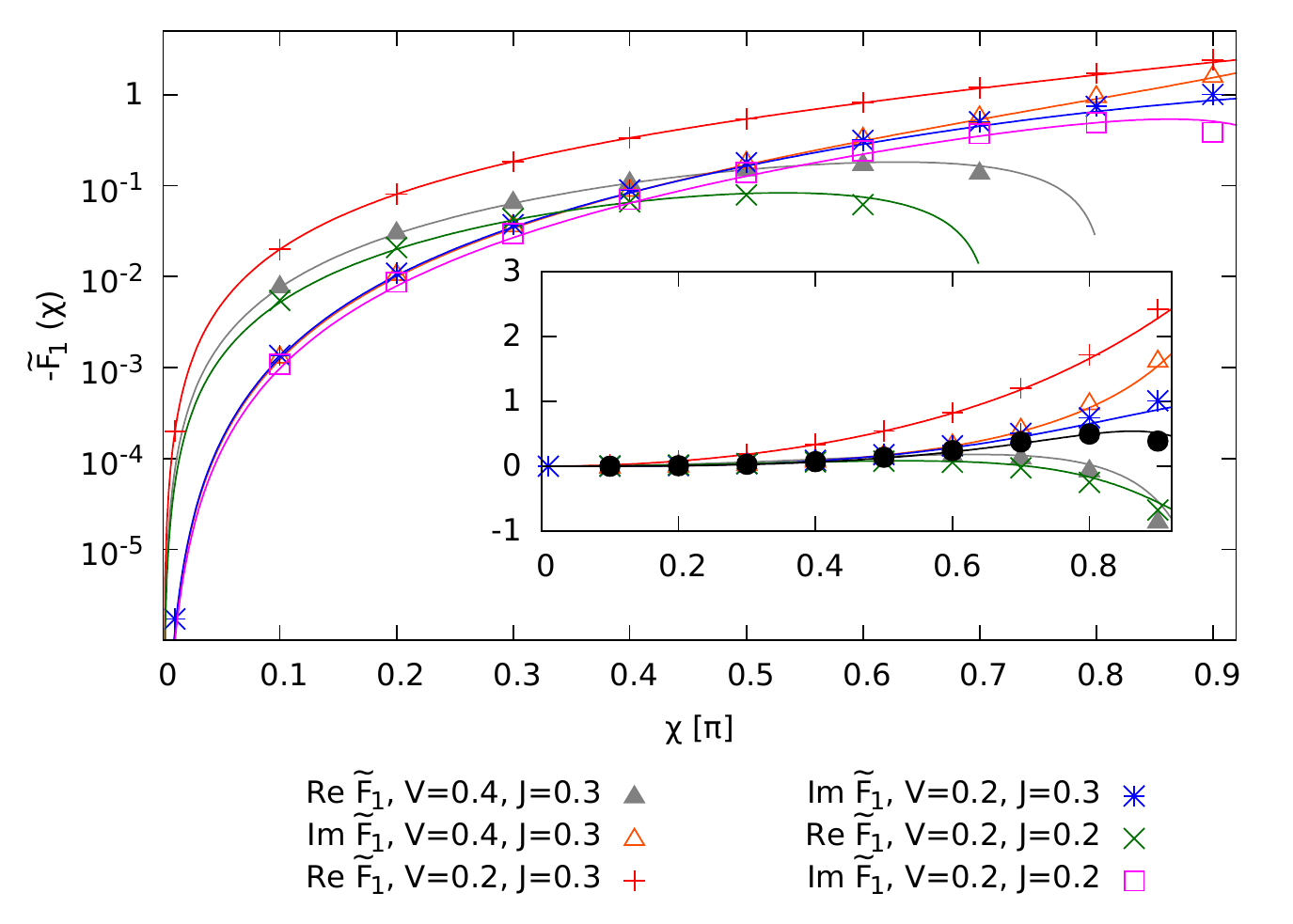}
\end{center}
\caption{Leading (left) and subleading (right) contributions to the full counting statistics of the non-interacting RLM (The inset gives the same result on a linear y axis.). Symbols correspond to the result obtained
   by fitting the series \eqref{eq:series} to the time evolution in numerical simulation, as shown in Fig.~\ref{fig:RLM-time-evolution}. The lines give the analytical results given by Eqs.~\eqref{eq:F0-RLM} and \eqref{eq:F1-RLM}.}\label{fig:dFxi}
\end{figure*}

Having obtained analytic expressions \eqref{eq:F0-RLM} and \eqref{eq:F1-RLM} for the coefficients of the series \eqref{eq:series} governing the time evolution of the CGF for the RLM, we can now compare this series with the full time evolution as obtained through a real time numerical simulation of the model (see Appendix \ref{sec:Size} for a brief outline of the numerical method).  This will allow us to see how well the series \eqref{eq:series}, which only contains two terms, compares to real numerical data, and furthermore allows us to see what features the time dependence may exhibit beyond this series.  The advantage of doing this for a non-interacting system is that one may obtain numerical data to relatively long times, which is not possible for strongly correlated interacting systems.  This will be a useful benchmark from which to interpret interacting results presented later in these proceedings.

In Fig.~\ref{fig:RLM-time-evolution}, we plot the real and imaginary parts of the CGF $\dot{F}(\chi,t_m)$ as a function of inverse measuring time $1/t_m$ for various values of the counting field $\chi$ and fixed values of the bias voltage $V_\textrm{SD}=0.2$ and model parameters $J'=0.3$.  The hopping parameter on the leads $J$ gives an overall energy scale to the problem and is fixed to be $J=1$.  The data points are obtained from numerical simulation, while the solid lines are from the series \eqref{eq:series}.  We note that in the series, rather than using bare measuring time $t_m$, we have replaced this by the conformal time $d(t_m,M)$, Eq.~\eqref{eq:ConformalTime}, where $M$ is the system size.  The conformal time, introduced in Ref.~\cite{Schmitteckert-Carr-Saleur-2014} takes into account the leading finite \textit{size} effects, and is briefly discussed in Appendix \ref{sec:Size}.  

We also note that the imaginary part of $\dot{F}$ tends to be positive, while the real part tends to be negative; the reasons for this can be seen by looking at the leading terms of the cumulant expansion Eq.~\eqref{eq:CumulantExpansion}.  In order to make the plots as clear as possible, we therefore plot both the real and imaginary parts as positive in most of the graphs, meaning that we plot $-\mathrm{real}(\dot{F})$ rather than $\mathrm{real}(\dot{F})$.  The plotting convention is always clearly stated on the axes of each graph.

Looking at Fig.~\ref{fig:RLM-time-evolution}, one can make several observations.  The first observation is that the agreement between the numerical data and the series \eqref{eq:series} is excellent over a number of orders of magnitude.  It is clear from the plots that the long time asymptotic of $\dot{F}(\chi,t_m)$ is indeed captured by the analytic expressions.  

The second observation is that the finite measuring time corrections $\tilde{F}_1$ beyond the infinite measuring time limit $\tilde{F}_0$ are much more important for the real part of this particular model.  In particular, if one is interested in the real part of $\dot{F}$ and has numerical data only up to $t_m\sim 40$ which is typical time scales reached in numerics on interacting systems, then one must extrapolate over several orders of magnitude to obtain the infinite time value $\tilde{F}_0$.  We will come back to this point in more detail when we discuss interacting models.  It is also worth mentioning that although the contribution from $\tilde{F}_1$ is not visible on the scale of the plot of the imaginary part in Fig.~\ref{fig:RLM-time-evolution}, this contribution is there and agrees well with the analytic formula; this is demonstrated in Fig.~\ref{fig:dFxi}.

The third observation from Fig.~\ref{fig:RLM-time-evolution} is that at smaller times, and in particular for larger values of $\chi$, there appears to be oscillations on top of the basic series \eqref{eq:series}.  We believe that these (decaying) oscillations are due to the sudden quench of the system when the counting field is switched on, in other words they depend on details of how the stopwatch for the FCS is started and stopped and are therefore do not depend only on the physics of the system in question (i.e. they are not universal).  We stress however that this is currently a hypothesis; future work is needed to understand these oscillations.

From the numerical data involved in each of these plots, one may fit the series \eqref{eq:series} to the time evolution in order to numerically determine the parameters $\tilde{F}_0$ and $\tilde{F}_1$.  This is plotted in Fig.~\ref{fig:dFxi} and compared to the analytic expressions \eqref{eq:F0-RLM} and \eqref{eq:F1-RLM} to good agreement.

\section{Extension to interacting systems}
\label{sec:FCS-I}

In Ref.~\onlinecite{Schmitteckert-Carr-Saleur-2014} we presented a conjecture for the form of the subleading (in inverse measuring time) term in the time evolution of the CGF.  Here, we review this conjecture, and present the latest numerical evidence supporting it.

\subsection{Conjecture for form of $\ln(V_\textrm{SD}t_m)$ corrections in interacting systems}

The scattering formulation which is used in Eqs.~\eqref{eq:LL} and \eqref{eq:NI-F1} can not be used in interacting systems, as such systems are not fully characterised by a single-particle transmission amplitude $T(\epsilon)$.  One can however ask the question: suppose we have managed to calculate the leading term in the CGF, $\tilde{F}_0(\chi)$.  From this knowledge, can we derive the size of the subleading term $\tilde{F}_1(\chi)$?

By simple comparison of Eqs.~\eqref{eq:LL} and \eqref{eq:NI-F1}, we see that
\begin{equation}
\tilde{F}_1 = \frac{\alpha}{2 } \left[ \left( \frac{\partial \tilde{F}_0}{dV_S} \right)^2 + \left( \frac{\partial \tilde{F}_0}{dV_D} \right)^2 \right],
\label{eq:conjecture1}
\end{equation}
where $\alpha=1/\pi$.  This formulation makes no reference to the single-particle scattering problem and we \textit{conjecture} that this expression \eqref{eq:conjecture1} remains true when interactions are added to the problem.  We allow in the conjecture that the constant of proportionality $\alpha$ may depend on the interaction strength, although the numerical work indicates that at least for the interacting resonant level model in Section \ref{sec:SDIRLM}, the value $\alpha=1/\pi$ is unchanged from the non-interacting value.  We stress however that this result can only hold for single channel systems -- as soon as there is a sum over different transmission channels, no such simple formula can exist.

For convenience we limit ourselves to cases which have particle-hole symmetry; meaning that $\frac{\partial \tilde{F}_0}{dV_S}=\frac{\partial \tilde{F}_0}{dV_D}$.  In the example of the non-interacting RLM, which satisfies $T(\epsilon)=T(-\epsilon)$, if one then applies the source and drain biases symmetrically $V_S=-V_D=V_\mathrm{SD}/2$, the conjecture \eqref{eq:conjecture1} simplifies to become
\begin{equation}
\tilde{F}_1 = \alpha \left( \frac{\partial \tilde{F}_0}{dV_\mathrm{SD}} \right)^2.
\label{eq:conjecture2}
\end{equation}

As previously mentioned, in the non-interacting case the logarithmic term in the series \eqref{eq:series1} arises due to the discontinuity in the Fermi function at zero temperature, the above derivative formula indeed relates $\tilde{F}_1$ to the physics at the Fermi edges.  Perturbatively, one may then certainly imagine an effective Landau picture of quasi-particles valid near the Fermi-surface.  These would make their contribution to the overall quantities such as current and noise (in essence $\tilde{F}_0$), although the calculation of these quantities require knowledge of all of the states in the energy window between the two Fermi edges so could not be reliably obtained simply from knowledge of the quasi-particles at the Fermi surface.  On the other hand the physics around the Fermi edge is all that is needed in the determination of $\tilde{F}_1$, which gives some hope that \eqref{eq:conjecture2} may remain valid in interacting systems.

We emphasize however that this line of reasoning, though intuitively appealing, has no rigour, and \eqref{eq:conjecture2} remains a conjecture.  We can however give two pieces of evidence that it remains true.  The first comes from looking at the lowest cumulants, while the second piece of evidence comes from verifying numerically that it holds, at least for one strongly interacting model.

\subsection{Corrections to cumulants}

Let us expand $\tilde{F}_0(\chi)$ as
\begin{equation}
\tilde{F}_0(\chi) = iI\chi - S\frac{\chi^2}{2},
\label{eq:CumulantExpansion}
\end{equation}
where $I$ is the current and $S$ is the zero-frequency shot noise.  According to conjecture \eqref{eq:conjecture2}, one then should have the corrections (expanded to quadratic order in $\chi$) as
\begin{equation}
\tilde{F}_1(\chi) = -\alpha G^2 \chi^2,
\label{eq:quadratic}
\end{equation}
where $G=\partial I/\partial V_\mathrm{SD}$ is the conductance of the system.

The consequences of this are twofold.  Firstly, we see that there are no (leading order) corrections from finite measuring time to the current.  Secondly, we see that the corrections to noise due to finite measuring time are
\begin{equation}
\Delta S \propto G^2.
\label{eq:NoiseFromConjecture}
\end{equation}
The significance of this relates to the frequency dependence of noise, which in non-interacting systems is known to be
\begin{equation}
S(\omega)-S(0) \propto G^2|\omega|.
\label{eq:NoiseFrequency}
\end{equation}
In \cite{Chamon-Freed-1999} it was shown that this relationship \eqref{eq:NoiseFrequency} remains true to all orders in perturbation theory when interactions are added, at least in a wide class of models, although the constant of proportionality $\alpha$ may depend on interaction strength.  Other work \cite{Lesage-Saleur-1997} has suggested that the relationship \eqref{eq:NoiseFrequency} may break down in the nonperturbative regime; however recent numerical work \cite{Branschaedel_Boulat_Saleur_Schmitteckert:PRL2010} has verified that this relation does indeed hold.  A full analytic understanding of this is still lacking however.

If one understands that a finite measuring time $t_m$ means that noise cannot be probed at zero frequency, and instead is measured at a characteristic frequency $1/t_m$, one sees that relations \eqref{eq:NoiseFrequency} and the conclusion from our conjecture \eqref{eq:NoiseFromConjecture} become equivalent.  This means that our conjecture is intimately related to the validity of the expression \eqref{eq:NoiseFrequency} in the presence of interactions.  While an analytic proof of this is still an open question, the existing literature is in support of this.

We also note that a similar relation between the measuring time dependence and the frequency dependence of the third cumulant has been previously discussed in non-interacting systems \cite{Salo-2006}.  However there is very little work on the frequency dependence of the third cumulant in the presence of interactions; this also remains an interesting direction for future study.

Having discussed the conjecture  in the context of the low cumulants, we now turn to the full CGF and show that \eqref{eq:conjecture2} holds for one interacting model.

\subsection{Application to the self-dual interacting resonant level model}
\label{sec:SDIRLM}

The interacting resonant level model is the same as the resonant level model, \eqref{eq:RLM} with the addition of an interaction $U$ between the resonant level and the leads.  The model is schematically represented in Fig.~\ref{fig:IRLM_lattice} and the Hamiltonian is given by:
\begin{equation}
{\cal H}_\mathrm{IRLM} = {\cal H}_\mathrm{RLM} + U\sum_{a=L,R} \left( \hat{d}^\dagger \hat{d} - \frac{1}{2} \right) \left( \hat{c}_{a,0}^\dagger \hat{c}_{a,0} - \frac{1}{2} \right)
\label{eq:IRLM}
\end{equation}
The quartic fermionic term here is the interaction, while the subtractions of $1/2$ is convenient to retain particle hole symmetry.

It turns out that for a certain value $U=2J$, this model has a certain self-duality \cite{Schiller-Andrei-2007} that allows the model to be solved exactly via Bethe ansatz, even out of equilibrium \cite{Boulat-Saleur-2008,Boulat-Saleur-Schmitteckert-2008,Branschaedel_Boulat_Saleur_Schmitteckert:PRL2010,CBS-2011}

The analytic expression for the CGF of the self-dual interacting resonant level model, as derived from thermodynamic Bethe ansatz, is given by \cite{CBS-2011}:
\begin{widetext}
\begin{equation}
\tilde{F}_0(\chi) = 
\begin{cases}
 i V_{\mathrm{SD}} \chi + V_{\mathrm{SD}}\!\!\sum_{m > 0}  \frac{a_4(m)}{2m} \left( \frac{V_{\mathrm{SD}}}{T_B'} \right)^{6m} \left( e^{-2\im m\chi} -1 \right), & V_\mathrm{SD}<V_c, \\
 V_{\mathrm{SD}}\!\!\sum_{m > 0}  \frac{2a_{1/4}(m)}{m} \left( \frac{V_{\mathrm{SD}}}{T_B'} \right)^{-3m/2} \left( e^{\im m \chi/2} -1 \right),    & V_\mathrm{SD} > V_c.
 \end{cases}
 \label{eq:F0-IRLM}
\end{equation}
\end{widetext}
Here
\begin{equation}
a_K(m)=\frac{(-1)^{m+1} \sqrt{\pi} \, \Gamma \left(1+Km \right) }{2 m! \,\Gamma \left(\frac{3}{2} +(K-1)m\right) },
\end{equation}
$T_B'\ = 2.7J(J'/J)^{4/3}$, and $V_c=\sqrt{3}T_B'/4^{2/3}$ is the radius of convergence of each series.   We note that unlike the non-interacting model, the interacting one cannot be solved exactly on the lattice, and the given result is only for the wide-band (field-theoretic) limit. The non-universal pre-factor $2.7$ relates the regularization of the field theory to the lattice model,  and is taken from earlier work \cite{Boulat-Saleur-Schmitteckert-2008} which studied the non-linear $IV$ characteristics.  This means that when we come to compare analytic and numerical results, there are no free fitting parameters and the plots simply show a direct comparison.  In order to do this however, parameters must be chosen carefully so that there is not a strong contribution from the finite band width in the numerical result.

We will return to the physical interpretation of the CGF \eqref{eq:F0-IRLM} in section \ref{sec:FCS-PT}; in this section we concentrate on the finite time corrections and the conjecture \eqref{eq:conjecture2}. Applying the conjecture, and furthermore taking the constant of proportionality $\alpha=1/\pi$ to be the same as the non-interacting value, one should then see the subleading term
\begin{widetext}
\begin{equation}
\tilde{F}_1(\chi) = 
\begin{cases}
\frac{1}{\pi} \left[ i \chi + \sum_{m > 0}  a_4(m)\frac{6m+1}{2m} \left( \frac{V_{\mathrm{SD}}}{T_B'} \right)^{6m} \left( e^{-2m\im\chi} -1 \right) \right]^2 & V_\mathrm{SD}<V_c, \\
\frac{1}{\pi} \left[  \sum_{m > 0}  a_{1/4}(m) \frac{1-3m/2}{m} \left( \frac{V_{\mathrm{SD}}}{T_B'} \right)^{-3m/2} \left( e^{\im m \chi/2} -1 \right) \right]^2 & V_\mathrm{SD} > V_c.
 \end{cases}
 \label{eq:F1-IRLM}
\end{equation}
\end{widetext}

\begin{figure*}
\begin{center}
\includegraphics[width=\columnwidth,clip=true]{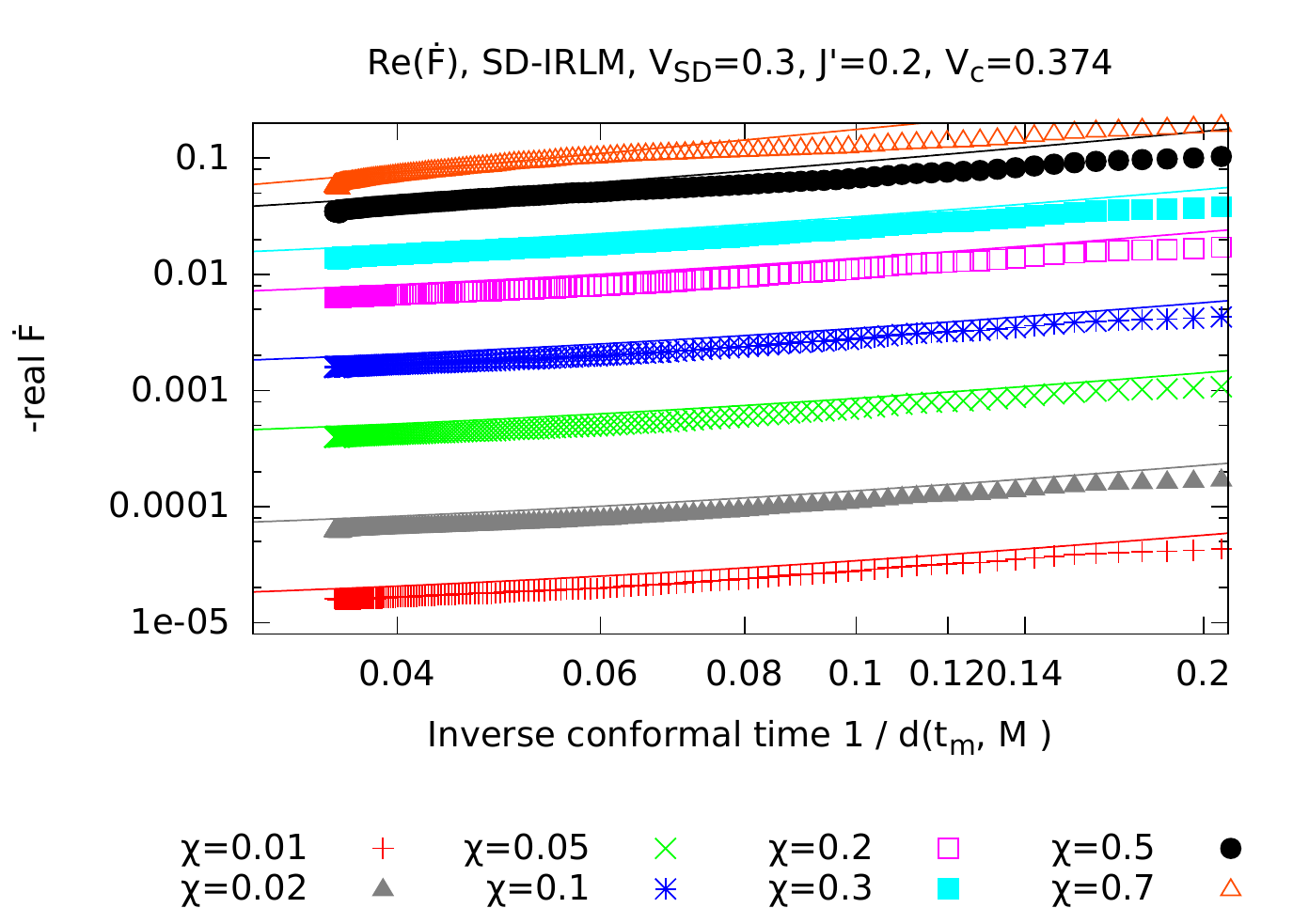}
\includegraphics[width=\columnwidth,clip=true]{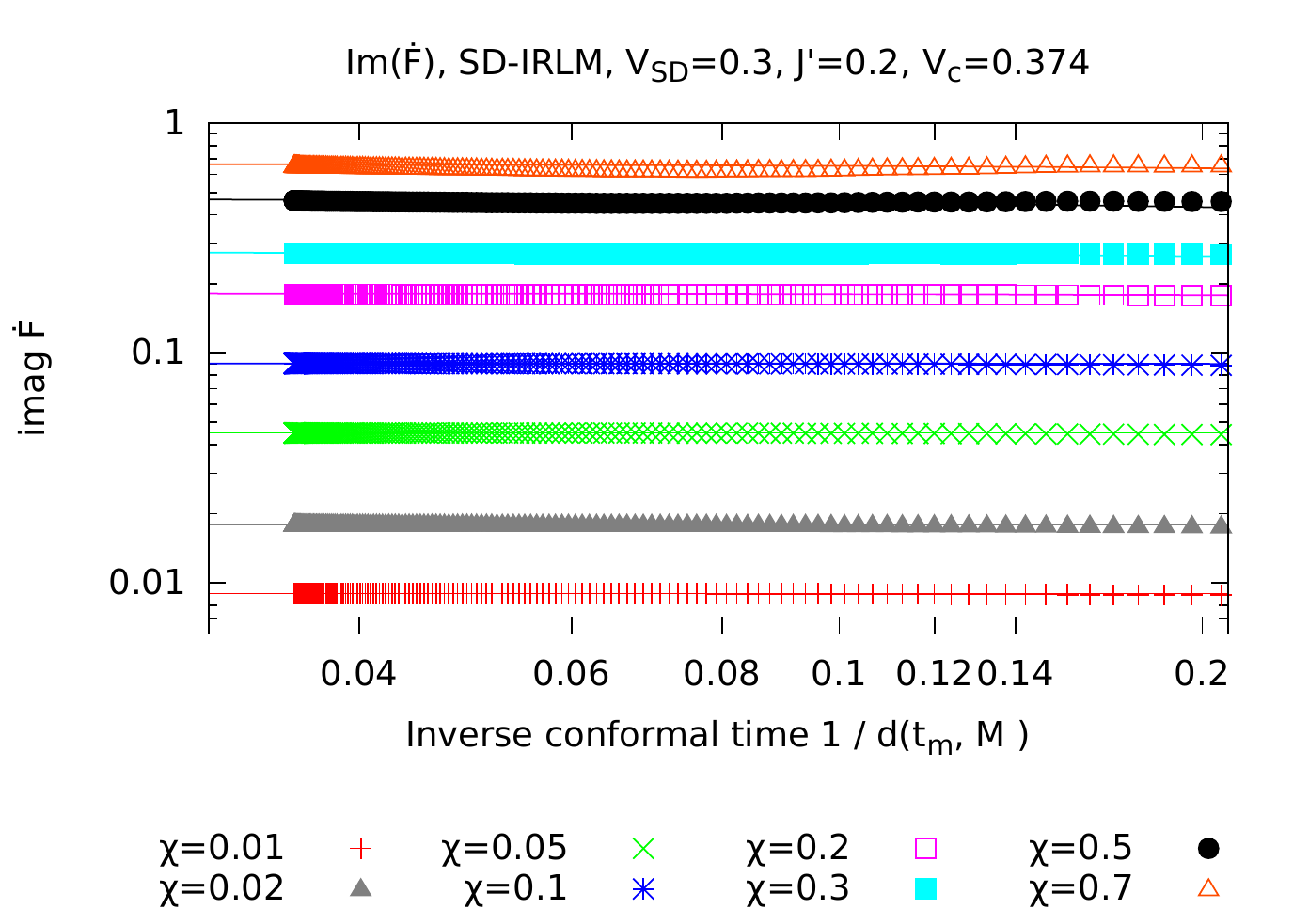}
\end{center}
\caption{An example of the measuring time evolution of the real (left panel) and imaginary (right panel) parts of $\dot{F}(\chi,t_m)$ for  the self-dual interacting resonant level model at $V_\textrm{SD}<V_c$.  The symbols come from a numerical evaluation with system size $M=240$.  The solid lines correspond to the series \eqref{eq:series} with the coefficients given in Eqs.~\eqref{eq:F0-IRLM} and \eqref{eq:F1-IRLM}, and with the conformal time substitution (see Appendix \ref{sec:Size}).}
\label{fig:SDIRLM-time}
\end{figure*}

\begin{figure*}
\begin{center}
\includegraphics[width=\columnwidth,clip=true]{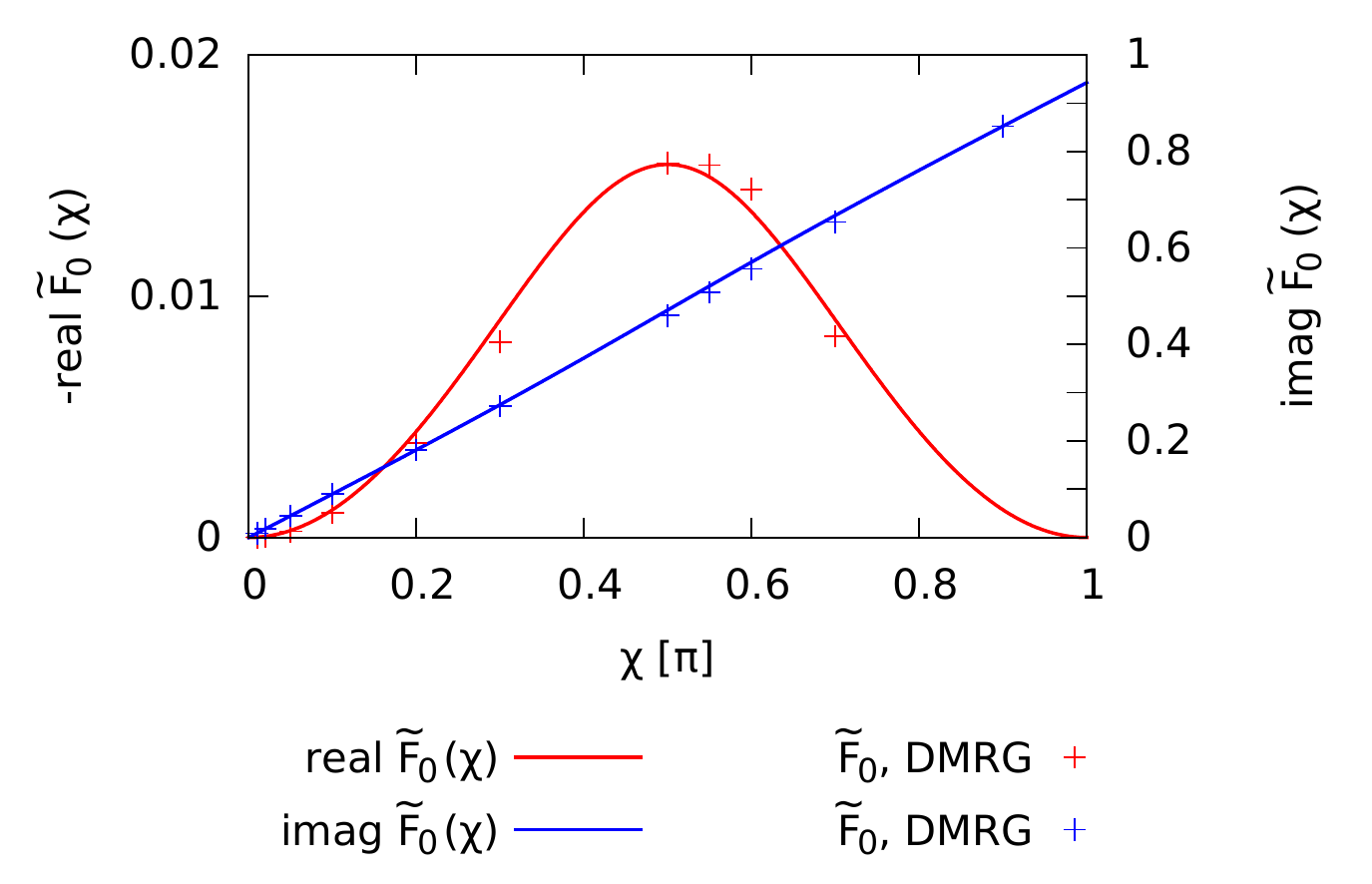}
\includegraphics[width=\columnwidth,clip=true]{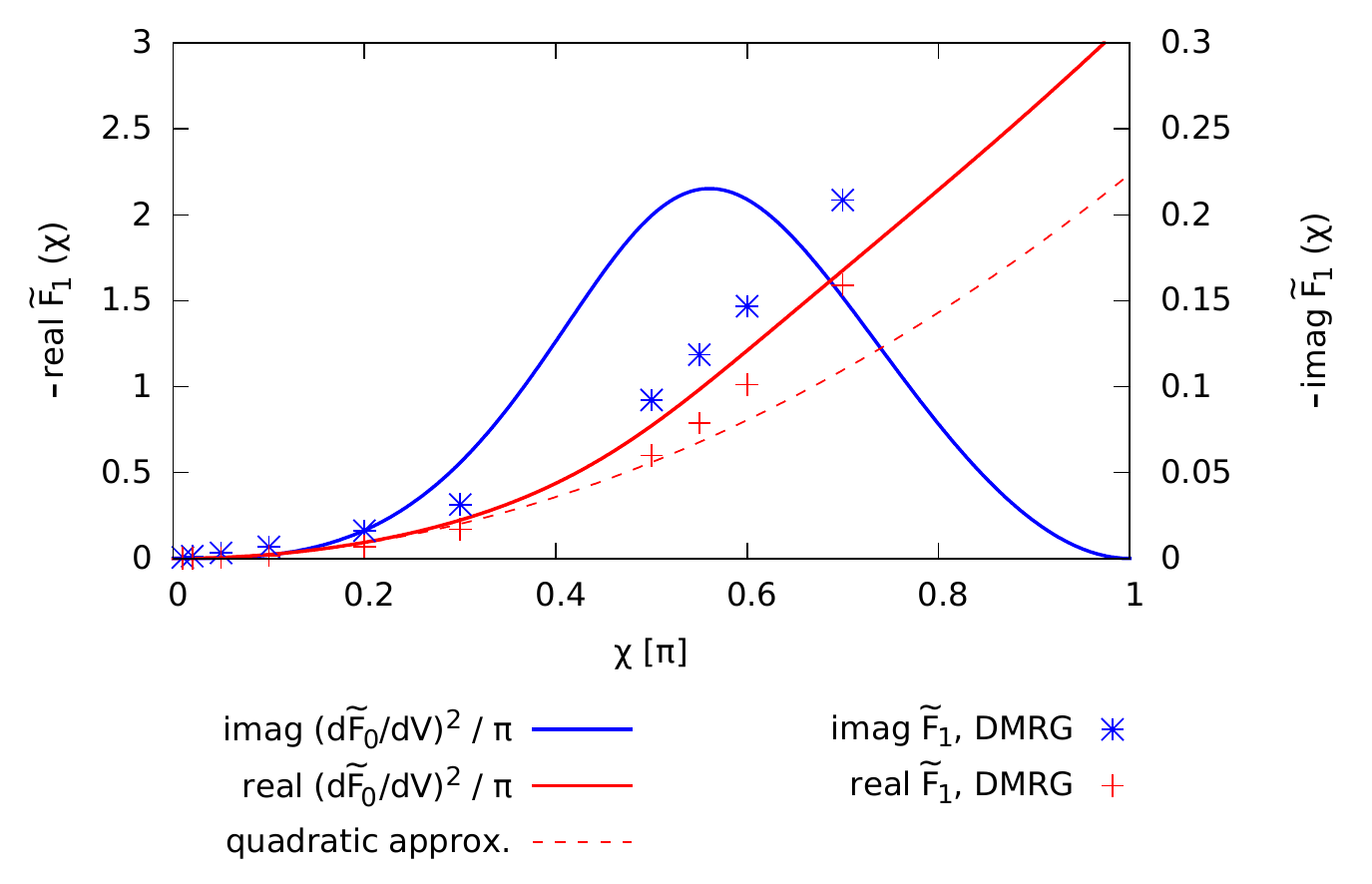}
\end{center}
\caption{The leading term $\tilde{F}_0$ (left) and subleading term $\tilde{F}_1$ (right) of the CGF for the interacting resonant level model at a small bias voltage $V_\mathrm{SD}<V_c$.  The symbols are obtained by fitting the series \eqref{eq:series} to the numerical data, while the solid lines are plots of the analytic result for $\tilde{F}_0$ in Eq.~\eqref{eq:F0-IRLM} and the  result for $\tilde{F}_1$ according to the conjecutre, \eqref{eq:F1-IRLM}.  In the real part of $\tilde{F}_1$, we also plot the quadratic approximation \eqref{eq:quadratic} to $\tilde{F}_1$ -- this clearly demonstrates that we see finite time corrections to cumulants higher than the second.}
\label{fig:SmallVoltage}
\end{figure*}

\begin{figure}
\begin{center}
\includegraphics[width=\columnwidth,clip=true]{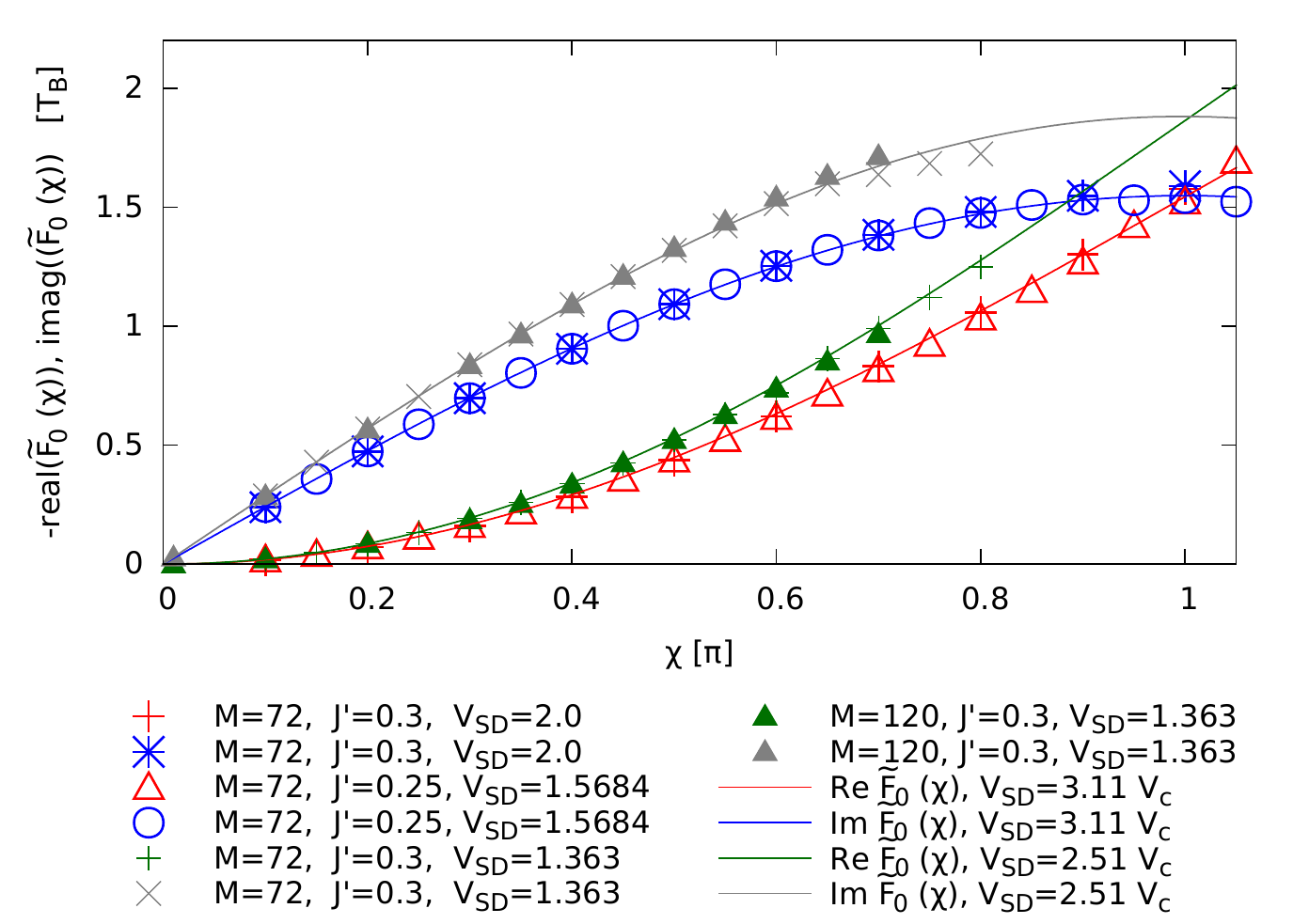}
\end{center}
\caption{The leading term $\tilde{F}_0$ of the CGF for the interacting resonant level model at different large bias voltages $V_\mathrm{SD}>V_c$.  The symbols give the numerical results, while the solid line follows the analytic result \eqref{eq:F0-IRLM}.  At these high biases, the subleading correction $\tilde{F}_1$ is too small to be reliably extracted from numerics.}
\label{fig:LargeVoltage}
\end{figure}

We are now in a position to compare these analytic predictions for $\tilde{F}_0$ and $\tilde{F}_1$ with results from real time numerical simulations -- see appendix \ref{sec:Size} for an overview of the numerical technique, as well as the two original papers \cite{CBS-2011,Schmitteckert-Carr-Saleur-2014} where these results were first presented.
 Numerically, one chooses a value of the counting field $\chi$, and time evolves the system to obtain $\dot{F}(\chi,t_m)$.  Fig.~\ref{fig:SDIRLM-time} shows some sample results of $\dot{F}$ as a function of measuring time $t_m$ for different values of $\chi$.  We mention here that a number of other plots of a similar nature are shown later in Fig.~\ref{fig:TimeSlices}.

In the plots of Fig.~\ref{fig:SDIRLM-time}, we also include the analytic result given by the series \eqref{eq:series} and the conjecture.  These appear to be in good agreement in the figure -- however one can go a stage further and extract the parameters $\tilde{F}_0$ and $\tilde{F}_1$ by fitting the series \eqref{eq:series} to the numerically obtained time evolution.  The results of this are shown in Figs.~\ref{fig:SmallVoltage} and \ref{fig:LargeVoltage}.

We begin by discussing the case $V_\mathrm{SD}<V_c$, which is shown in Fig.~\ref{fig:SmallVoltage}.  We begin by discussing the real part.  As seen in the figure, there is a very good agreement between the numerically determined $\tilde{F}_0$ and the analytic result.  The incredible thing here is that to obtain the numerical result, one must extrapolate $\dot{F}$ about an order of magnitude from the finite-time results available.  In other words, for the time scales that can be simulated on the computer, the result is dominated by the $\tilde{F}_1$ contribution -- one should carefully read the scale on the axis of the graphs of $\tilde{F}_0$ and $\tilde{F}_1$.  Looking at the result for the real part of $\tilde{F}_1$, we also see that the numerical data agrees well with the analytic result \eqref{eq:F1-IRLM}.  This  is strong numerical evidence supporting our conjecture that the formula \eqref{eq:conjecture2} does indeed hold in the presence of interactions.  In Fig.~\ref{fig:SmallVoltage}, we have also plotted the quadratic approximation $\tilde{F}_1 \propto \chi^2$, which is the finite time (frequency) correction to the shot noise that has been previously discussed in \cite{Branschaedel_Boulat_Saleur_Schmitteckert:PRL2010}.  From the plot, it is clear that the conjecture \eqref{eq:conjecture2} correctly captures the correction to cumulants beyond this.

We next look at the imaginary part of $\tilde{F}_{0,1}$, also plotted in Fig.~\ref{fig:SmallVoltage}.  Here it is clear that the expression for $\tilde{F}_0$ fits rather well, however it would be difficult to say the same for $\tilde{F}_1$.  There are several issues going on here though.  First, one must again look at the vertical scales of these graphs.  For the imaginary part of $\dot{F}$, the finite measuring time corrections $\tilde{F}_1$ are a very small correction on top of the long-time limit; as an order of magnitude this is $1\%$ of the signal at $t_m = 10$, and less as $t_m$ increases.  Combined with this, there are other contributions to $\dot{F}$ beyond the terms given in the series \eqref{eq:series}.  In particular, looking at one of the first panels of Fig.~\ref{fig:TimeSlices} that appears later, one see's that the imaginary part of $\dot{F}$ exhibits certain oscillations -- presumably of the same (as yet unknown) origin as those previously seen in the non-interacting case.  This oscillations make it difficult to fit the small overall slope of the numerical data reliably, as the answer depends on the exact range of data used for fitting.  Combining this with the small overall size of $\tilde{F}_1$ makes the numerical task daunting.  We emphasize that this is work in progress, and will be reported elsewhere when enough data is available to make reliable predictions; however with this in mind, the discrepancy between the analytic formal and the numerical data (remembering that there are no fitting parameters whatsoever in the analytic formula) is not as severe as it may look.

For completeness, we finally turn our attention to the high voltage regime, $V_\mathrm{SD}>V_c$, where the analytic result is given by the other series in Eqs.~\eqref{eq:F0-IRLM} and \eqref{eq:F1-IRLM}.  For these large bias voltages, the overall scale of $\tilde{F}_1$ as compared to $\tilde{F}_0$ is even worse than the previously discussed case and we therefore unfortunately cannot fit $\tilde{F}_1$ reliably to the numerical data.  Instead, we give a comparison between the numerical and analytic $\tilde{F}_0$ in Fig.~\ref{fig:LargeVoltage}.  It is clear there is a good agreement.

The advantage of the self-dual interacting resonant level model in testing the conjecture \eqref{eq:conjecture2} is that analytic results exist for the CGF, which may be easily differentiated and compared to numerics to either support or disprove the conjecture.  While we are not aware of exact results for any other models, certain things may be checked, see Appendix \ref{sec:otherU} where the latest results in this direction are presented.

\section{Non-equilibrium phase-transitions}
\label{sec:FCS-PT}

We now return to the analytic formula \eqref{eq:F0-IRLM} for the (long time limit of the) CGF for the self-dual interacting RLM, and discuss the physical interpretation.  We being with the small bias voltage case $V_\mathrm{SD} \ll V_c$.  The first term of Eq.~\eqref{eq:F0-IRLM}, $-iV_\textrm{SD} \chi$, corresponds to the CGF of a system with perfect transmission; in such a case there is no source of noise and all higher cumulants are zero.  The fact the `impurity' level is tuned to resonance makes it unsurprising that the transport (at low bias voltages) may be considered as a weak correction to perfect transmission.  The remaining terms in the series give a power series in $(V_\mathrm{SD}/T_B')^6$, and may be interpreted as the weak backscattering current, and it's associated statistics.  Each term is proportional to $(e^{-2mi\chi}-1)$ which is the CGF for a Poissonian process moving charges of $2m$ from the right to the left (due to the minus sign -- this validates the interpretation as the backscattering correction to perfect transmission) \cite{Saleur-Weiss-2001,Komnik-2007,CBS-2011}.  The constants aren't all positive, due to quantum interference between the different processes.  The important point however is that the backscattered charges are all multiples of $2$, or in other words the quasi-particles that may be backscattered have charge $2e$ if units are resurrected.

In the opposite limit $V_\mathrm{SD}\gg V_c$, one notices that the periodicity of $\tilde{F}_0$ is different -- this time the sum may be interpreted as a sum of Poissonian processes carrying charges $m/2$ from left to right; or in other words the charge on the quasi-particles that may be transmitted in this limit is $e/2$ (restoring units).

This fractionalisation, by which the fundamental charge changes from $2$ at low bias to $1/2$ at high bias was first observed by studying the Fano factor of the shot noise \cite{Boulat-Saleur-Schmitteckert-2008}.  The shot noise however gave no indication of how this fractionalisation occurs.  From studying the FCS \eqref{eq:F0-IRLM} however, we will be able to interpret this as a bifurcation of the CGF at $V=V_c$; in some sense we will show that $V=V_c$ is the point of a non-equilibrium phase transition \cite{CBS-2011}.

\subsection{Bifurcation in the CGF of the self-dual IRLM}
\label{sec:bif}

\begin{figure}
\begin{center}
\includegraphics[width=\columnwidth,clip=true]{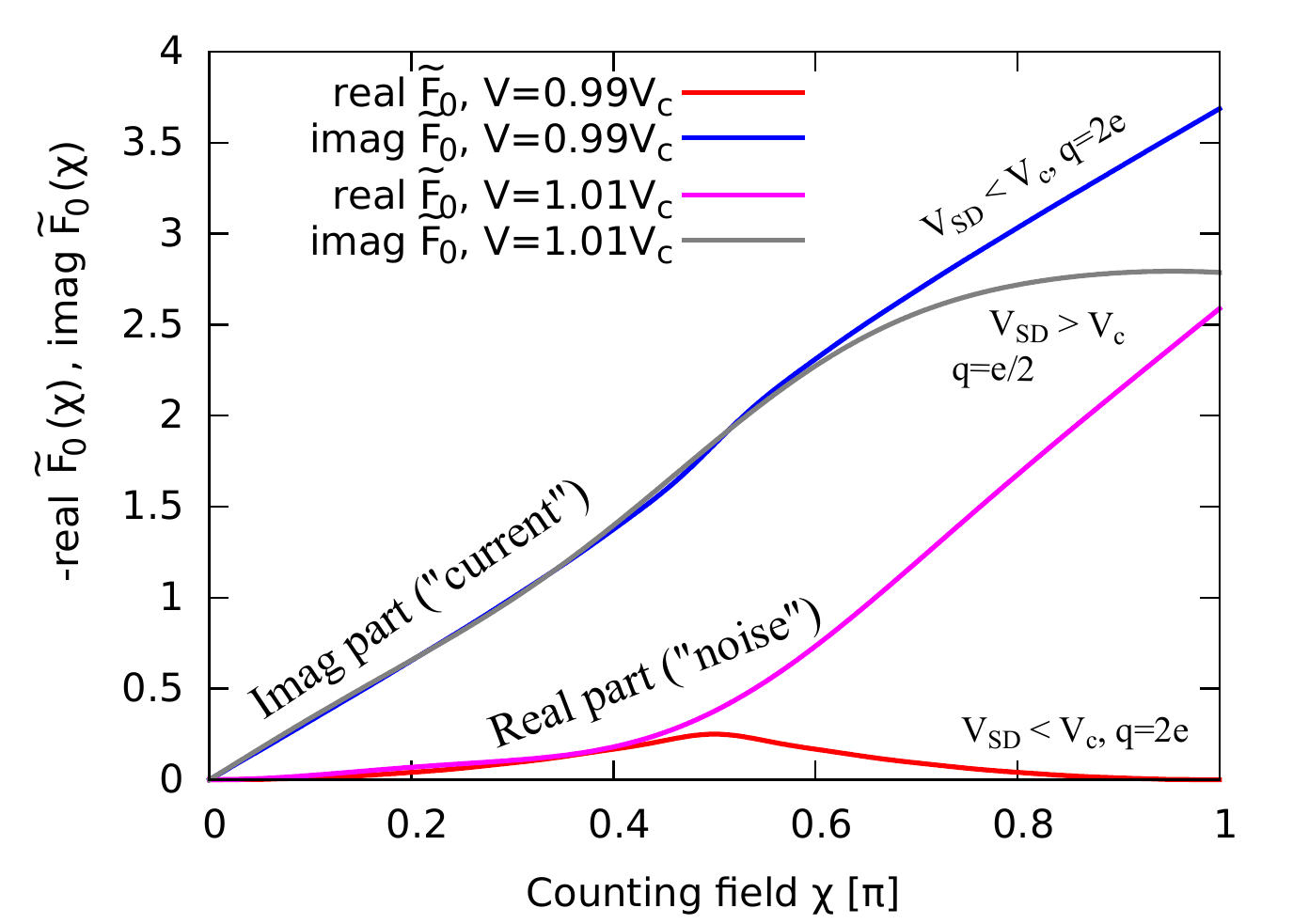}
\end{center}
\caption{A plot showing the analytic result \eqref{eq:F0-IRLM} for the long time limit $\tilde{F}_0$ of the CGF of the self-dual interacting RLM just above and just below the critical voltage $V_c$.   This plot demonstrates the bifurcation at $\chi=\pi/2$ which may be interpreted as a non-equilibrium phase-transition.}
\label{fig:bifurcation}
\end{figure}

While the series representation of the CGF, \eqref{eq:F0-IRLM} is useful away from $V=V_c$, at $V\sim V_c$ all terms of the series become important and the the behavior of $\tilde{F}_0(\chi)$ is mathematically opaque.  Fortunately, one may recognise that the series for high and low voltages are actually hypergeometric and both give different series representations of the same (generalised) hypergeometric function.  While one may do this for $\tilde{F}_0$ itself, it is far more convenient to take a derivative and study the counting current $I(\chi)=i\partial \tilde{F}_0/\partial \chi$.  Carrying out the summation gives \cite{CBS-2011}
\begin{equation}
I(\chi) = V_\mathrm{SD} \; {}_3F_2\left( \left\{ \frac{1}{4},\frac{3}{4},1\right\}, \left\{ \frac{5}{6},\frac{7}{6} \right\}, -z^2 \right),
\end{equation}
where $z=(V_\mathrm{SD}/V_c)^3e^{-i\chi}$.  

While this representation as a rather exotic hypergeometric function, ${}_3F_2$, is not useful in terms of numerical evaluation, the analytic properties of hypergeometric functions are well documented (see, e.g. \cite{AS}).  The function above ${}_3F_2(\cdots, -z^2)$ has two branch points at $z=\pm i$, with branch cuts stretching from these points to $\pm i\infty$.  This indicates that for $|z|<1$, i.e. for $V_\mathrm{SD}<v_c$, the function has a single branch, and from the definition of $z$ we see that $I(\chi)$ is $\pi$ periodic in $\chi$, giving the fundamental charge of $2$ previously mentioned.  On the other hand, if $|z|>1$, i.e. $V_\mathrm{SD}>V_c$, the function crosses the branch cuts at $\chi=\pi/2,3\pi/2$.  By choosing $I(\chi)$ to be continuous as a function of $\chi$, this implies that $I(\chi)$ must now be $4\pi$ periodic, which in turn implies a fundamental charge of $1/2$.

We now see that the fractionalisation of charge happens discontinuously at $V_\mathrm{SD}=V_c$; the branch point leading to a bifurcation in the CGF.  This can be seen in Fig.~\ref{fig:bifurcation}, where the CGF $\tilde{F}_0(\chi)$ is plotted for $V_\mathrm{SD}$ slightly above and slightly below $V_c$.  It is worth pointing out however that while this branch cut leads to a discontinuity in the CGF as a function of $V_\mathrm{SD}$, this discontinuity occurs at $\chi>\pi/2$.  All of the cumulants, which are given as derivates around $\chi=0$, are smooth analytic functions of $V_\mathrm{SD}$ at $V_\mathrm{SD}=V_c$.  In other words, one needs all of the cumulants, i.e. the \textit{full} counting statistics in order to see and understand this transition.

\subsection{Effect of finite-time}

\begin{figure*}
\begin{center}
\includegraphics[width=0.49\textwidth,clip=true]{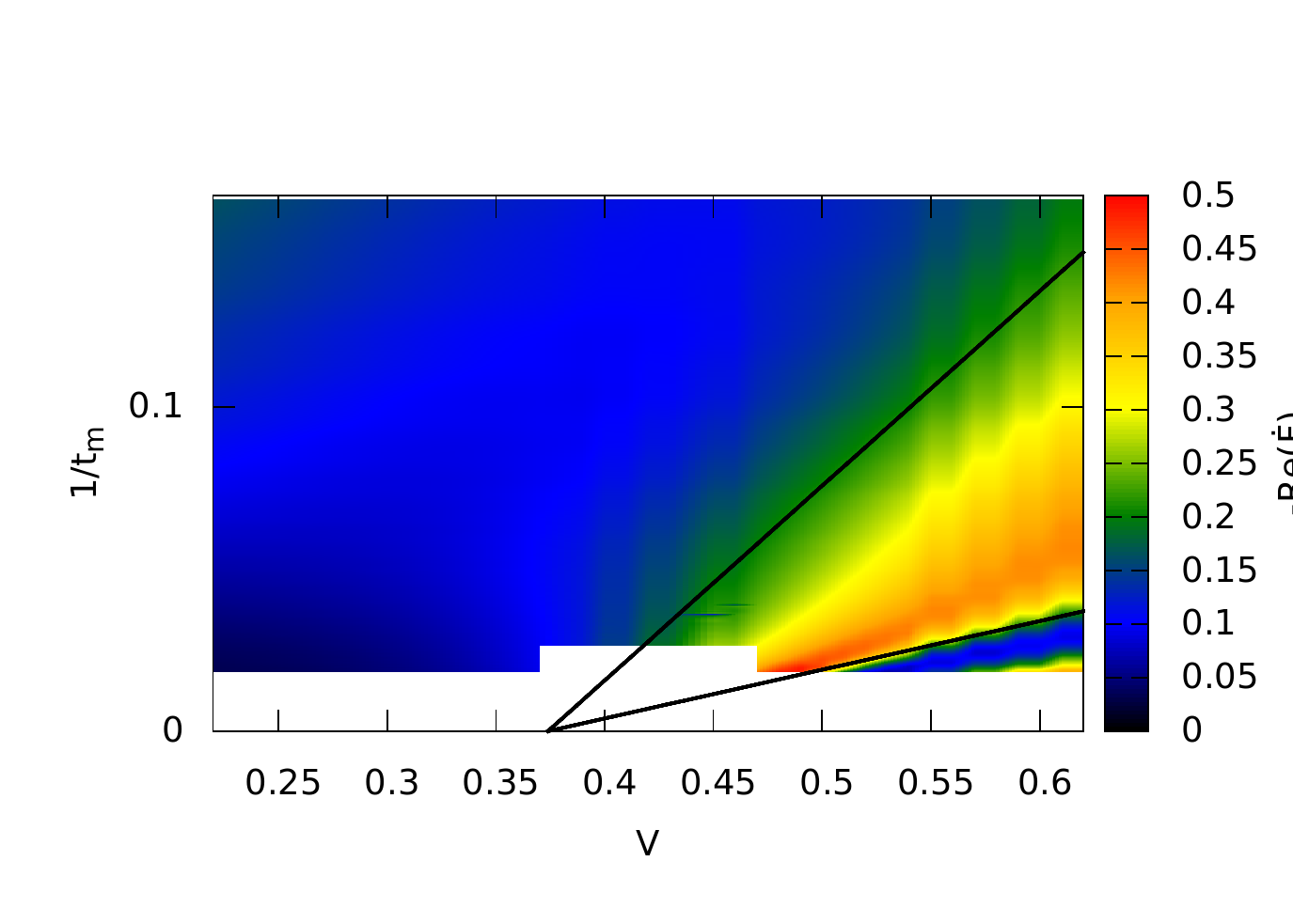}
\includegraphics[width=0.49\textwidth,clip=true]{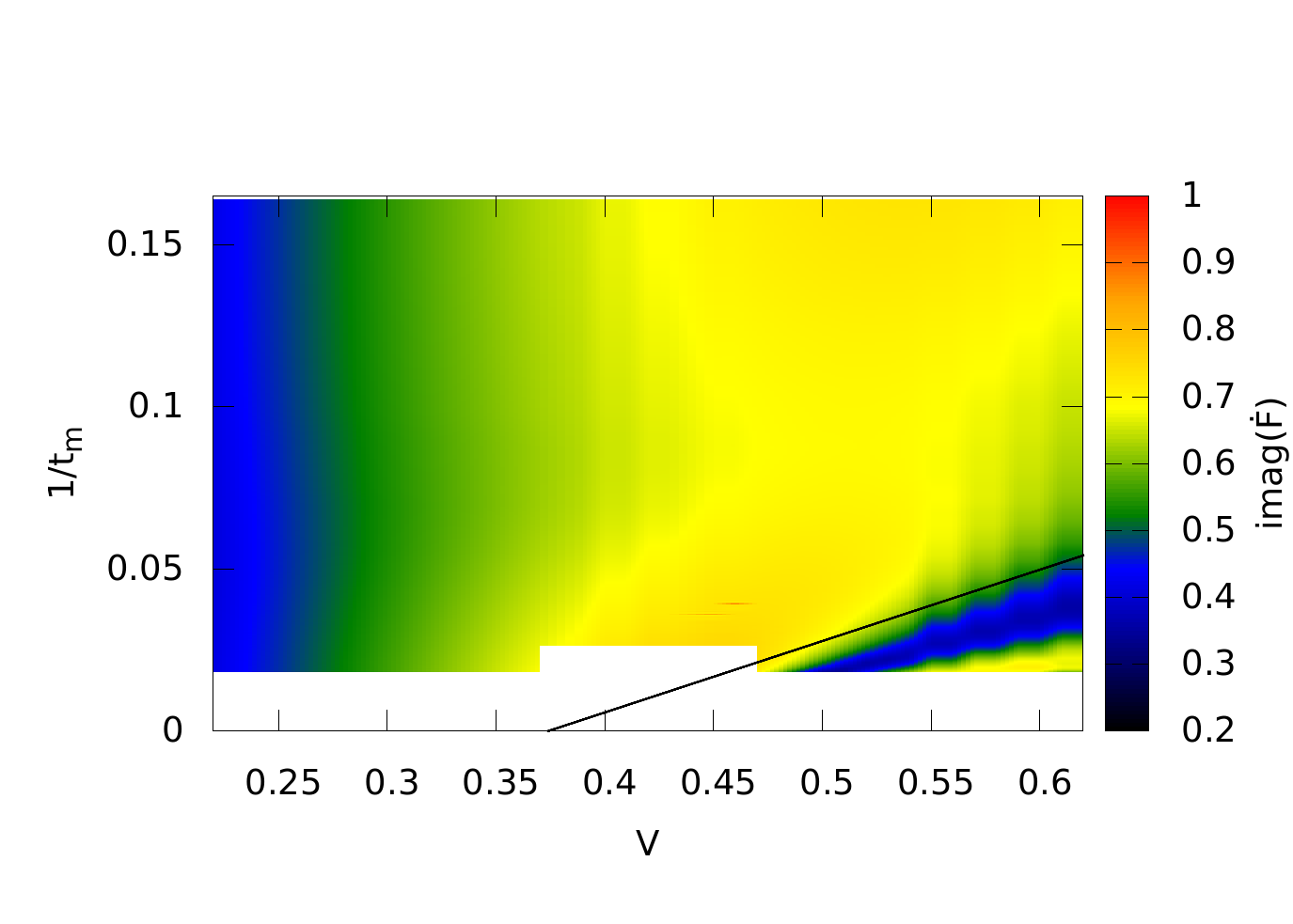}
\end{center}
\caption{A density plot of the real (left) and imaginary (right) part of the CGF $\dot{F}$ of the interacting RLM as a function of measuring time $t_m$ and bias voltage $V_\mathrm{SD}$ for $\chi=0.6\pi$.  For the model parameters chosen, the critical bias voltage $V_c=0.374$.  The black lines satisfy $1/t_m \propto (V_\mathrm{SD}-V_c)$ and are intended as a guide to the eye, indicating that the proximity to the phase transition is having a clear and important effect on the time evolution of the CGF.}
\label{fig:coolplot}
\end{figure*}

The previous subsection analysed only the (infinitely) long time limit, $\tilde{F}_0$, and found a sharp transition due to singularities (branch points) in the CGF.  We would ask: what is the effect of finite-time on this transition?  There are many important aspects of this question, including:
\begin{enumerate}
\item In analogy to finite size on equilibrium phase transitions, is the singularity present at finite time, or is the dependence on $V_\mathrm{SD}$ smooth at any finite time becoming singular only in the $t_m\rightarrow \infty$ limit?
\item Does finite time play a role in our ability to see fractional charges?  In other words, does the $2\pi$ periodicity of the CGF remain at all finite times, becoming larger only in the infinite time limit?
\end{enumerate}
While we are not yet able to answer these questions, we now present results about the measuring time dependence of the CGF near the critical voltage $V_c$ that will give a start in understanding these points.

It is clear that the conjecture \eqref{eq:conjecture2} relies on $\tilde{F}_0$ being a smooth function of $V_\mathrm{SD}$; near the critical bias voltage $V_c$, this is no longer the case.  This means that in the vicinity of $V_c$, the conjecture cannot hold, or at the very minimum there must be a lot of physics involved in the measuring time dependence of $\dot{F}$ that is not captured by the simple series \eqref{eq:series} and the conjecture \eqref{eq:conjecture2}.  In order to study this numerically, we take a fixed value of $\chi>\pi/2$ (we choose $\chi=0.6\pi$) and study $\dot{F}$ as a function of measuring time $t_m$ and bias voltage $V_\mathrm{SD}$ chosen such that $V_\mathrm{SD} \approx V_c$.

The results are shown as a density plot in Fig.~\ref{fig:coolplot}.  For the model parameters chosen ($J'/J=0.2$), one has $V_c \approx 0.374$.  Superimposed on the plots are various lines $1/t_m \propto (V_\mathrm{SD}-V_c)$, which show that the time evolution of the CGF has distinctive features due to the proximity to the critical point.  We would like to point out that there is an analogy between our data in Fig.~\ref{fig:coolplot} and the standard view of (equilibrium) quantum critical points \cite{Sachdev}, where inverse measuring time in our case takes the role usually played by temperature.  Whether or not this analogy is a useful way of thinking of the non-equilibrium problem however is an open question.

\begin{figure*}
\begin{center}
\includegraphics[width=0.49\textwidth,clip=true]{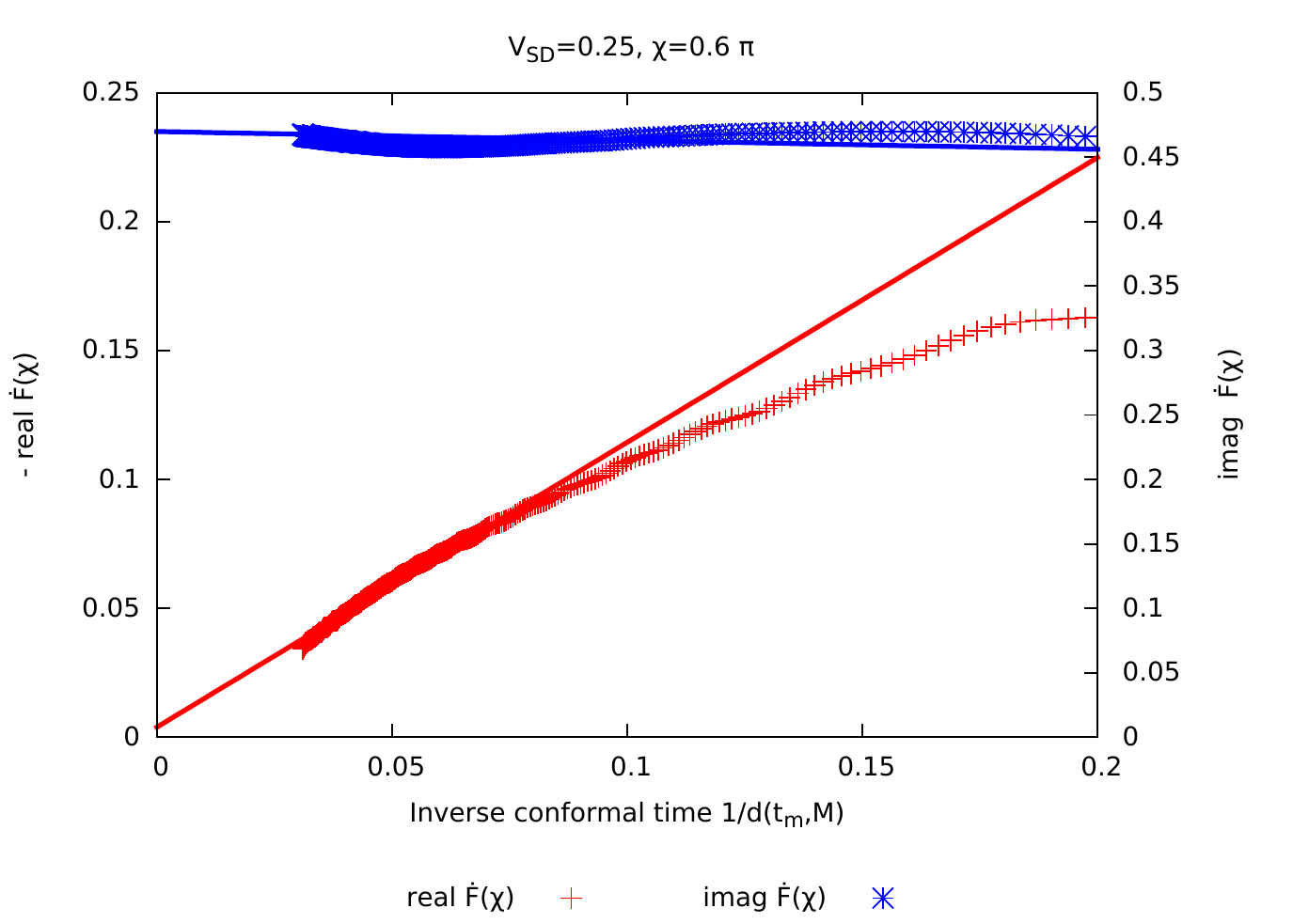}
\includegraphics[width=0.49\textwidth,clip=true]{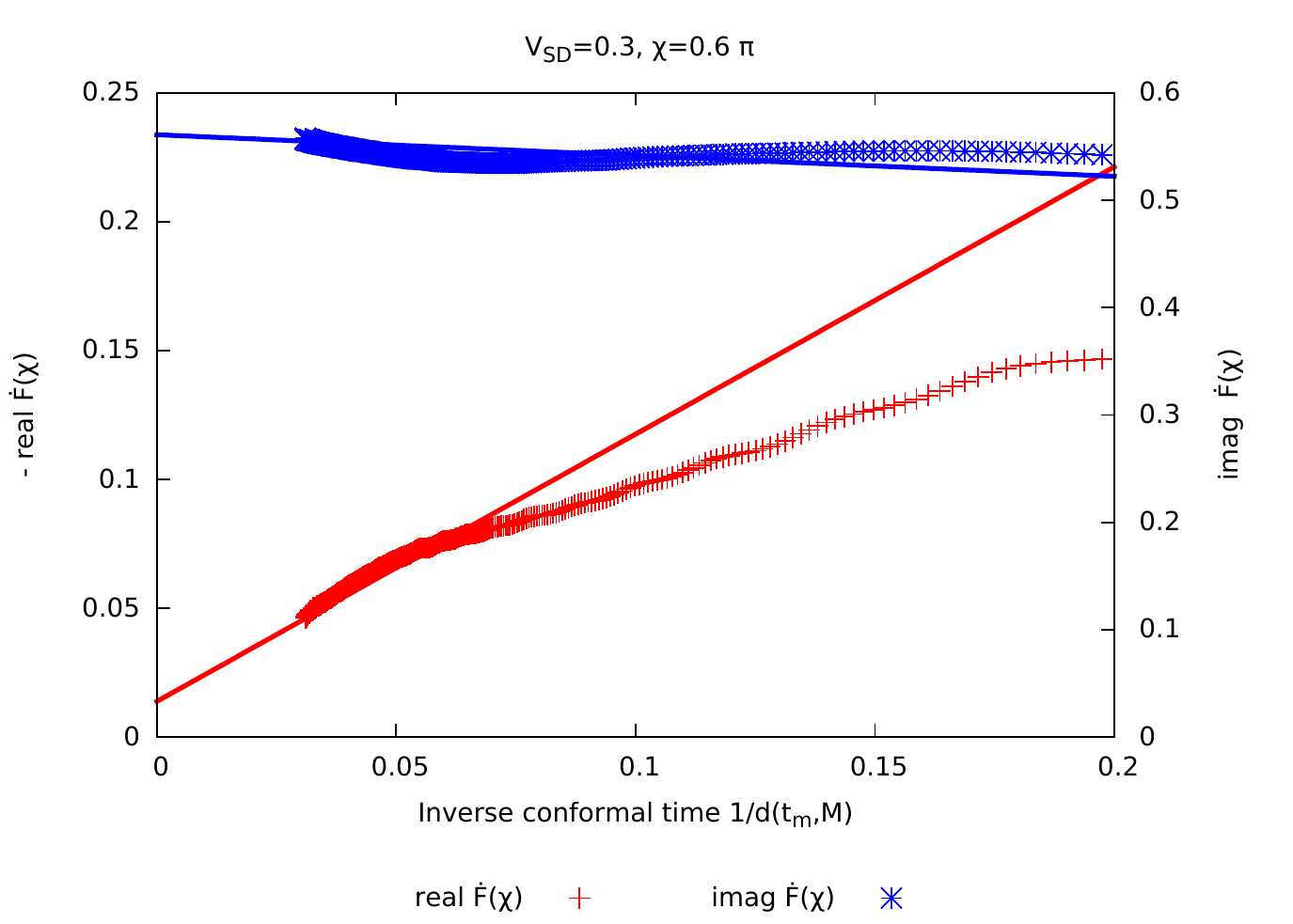}
\includegraphics[width=0.49\textwidth,clip=true]{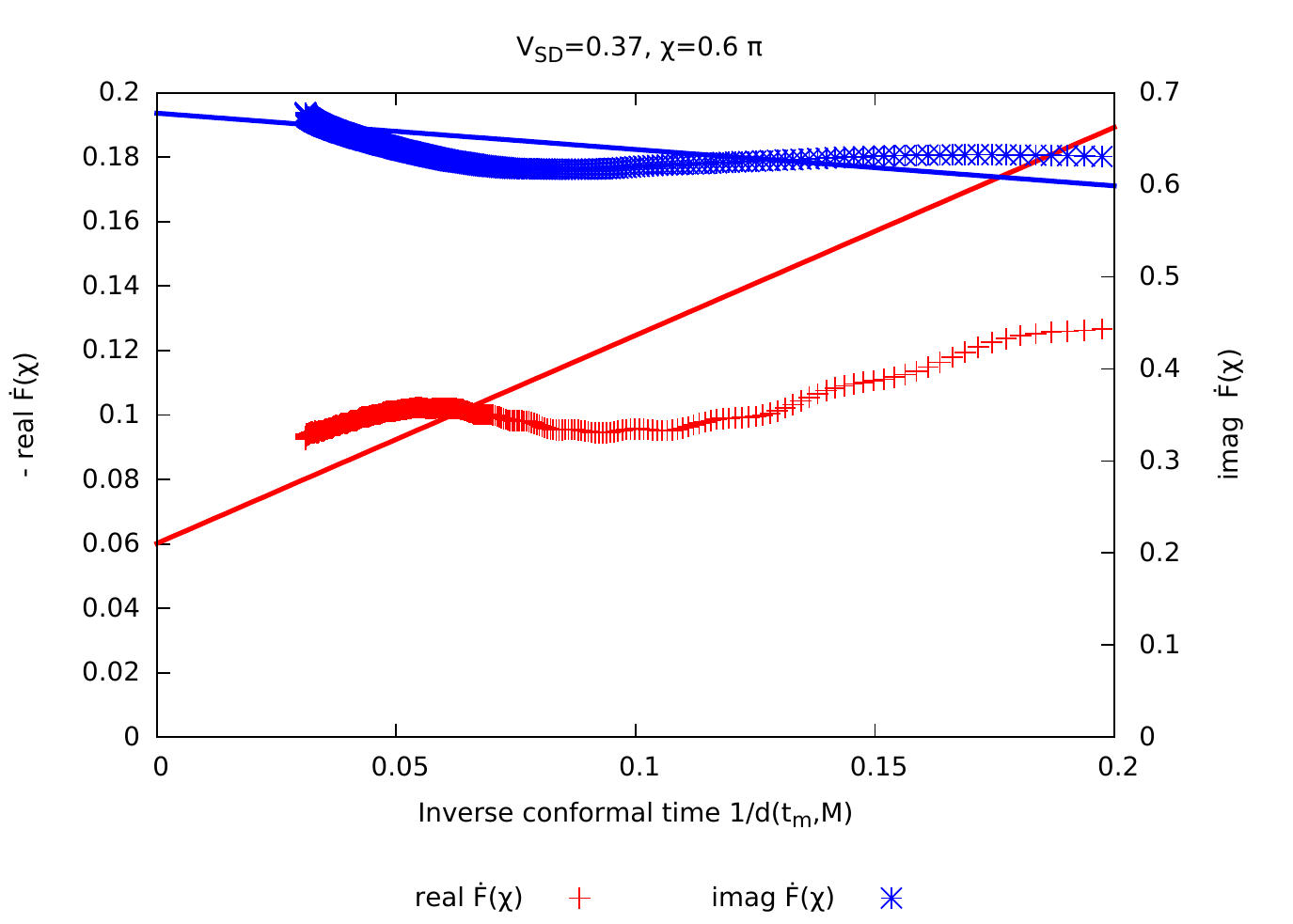}
\includegraphics[width=0.49\textwidth,clip=true]{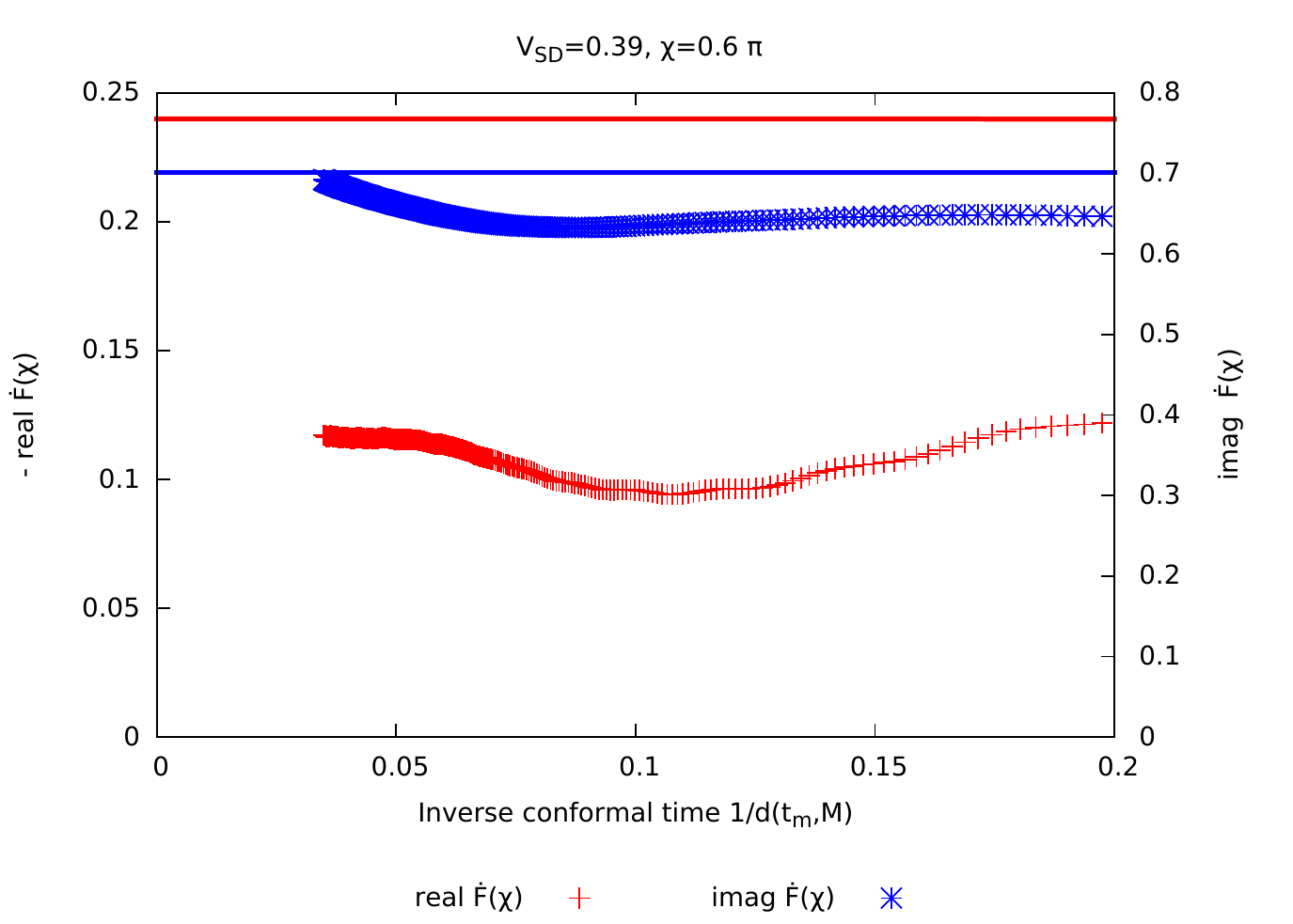}
\includegraphics[width=0.49\textwidth,clip=true]{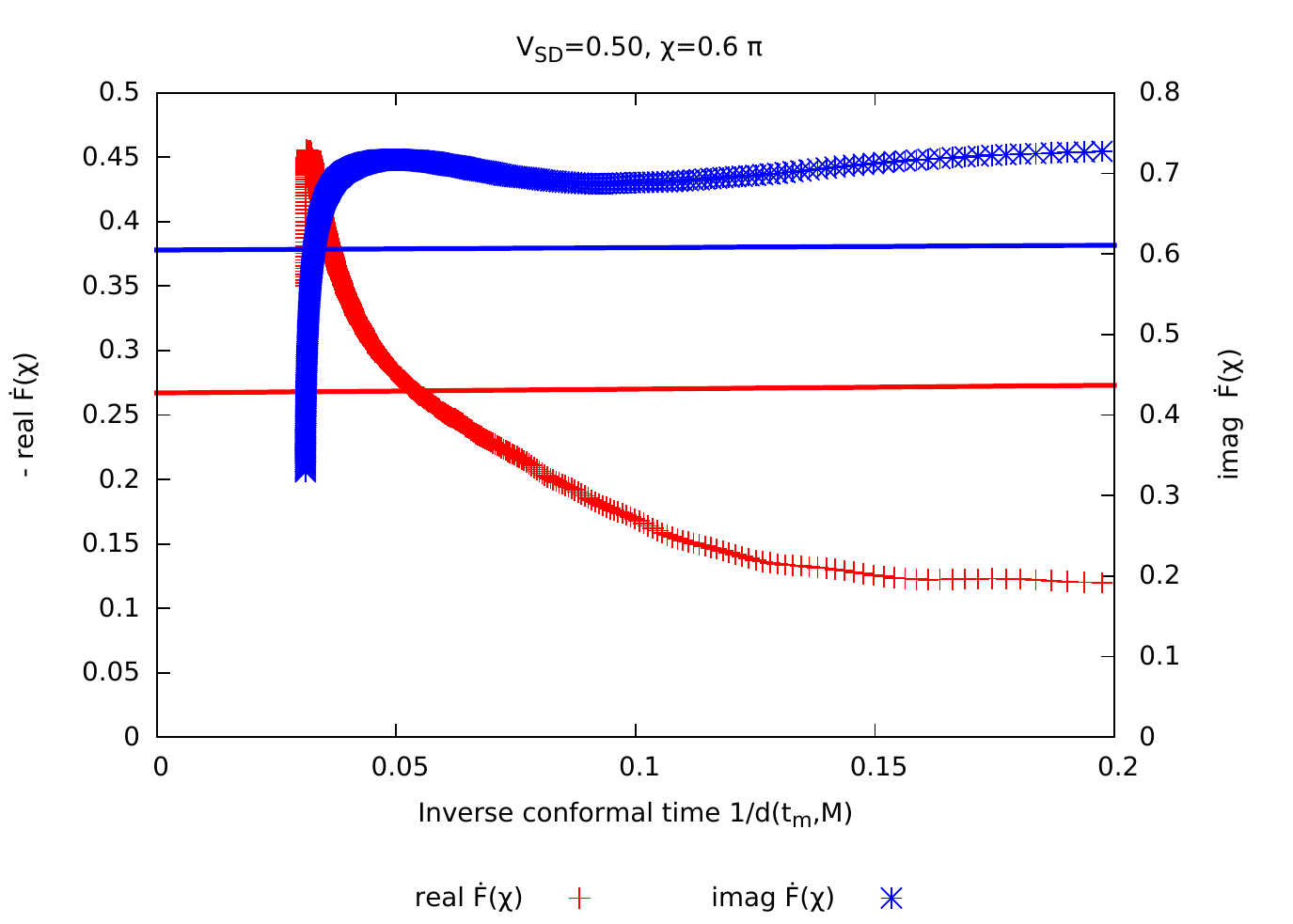}
\includegraphics[width=0.49\textwidth,clip=true]{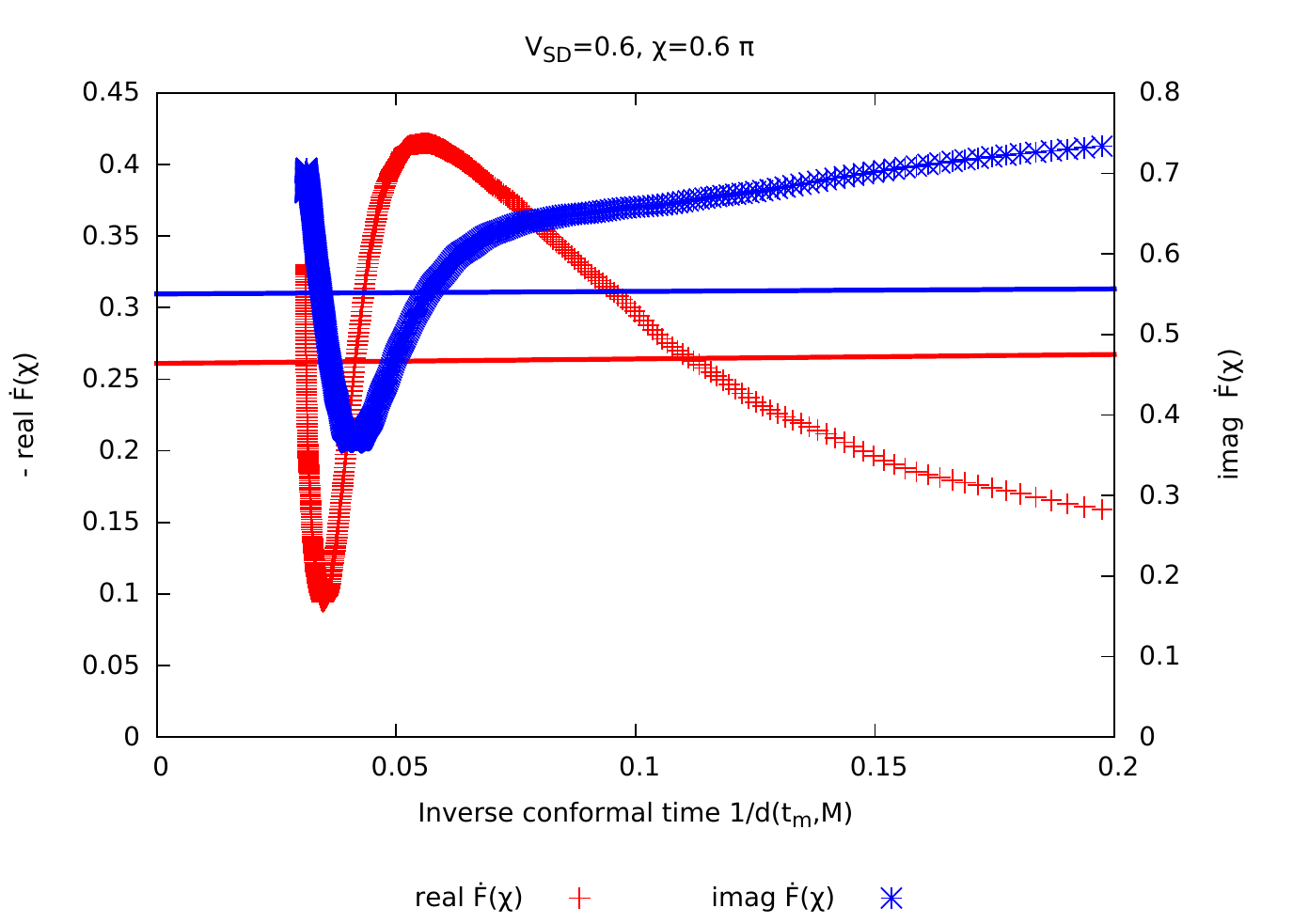}
\end{center}
\caption{Time evolution of $\dot{F}$ for various values of $V_\mathrm{SD}$ close to the critical value.  These plots are vertical cross sections of the density plot shown in Fig.~\ref{fig:coolplot}.  In addition to the numerical data plotted as symbols, the analytic results $\dot{F}=\tilde{F}_0+\tilde{F}_1/t_m$ according to the series \eqref{eq:series} and conjecture \eqref{eq:conjecture2} are plotted as solid lines for comparison.}
\label{fig:TimeSlices}
\end{figure*}

To investigate further the features in the time evolution that are indicated in Fig.~\ref{fig:coolplot}, we take some vertical cross sections of this plot, in other words we plot $\dot{F}$ as a function of $t_m$ for a selection of values of $V_\mathrm{SD}$ close to the critical value.  This is shown in Fig.~\ref{fig:TimeSlices}, where the analytic result according to the series \eqref{eq:series} and conjecture \eqref{eq:conjecture2} are also plotted for comparison.  We start with the first two graphs, which are for $V_\mathrm{SD}<V_c$.  The imaginary part of $\dot{F}$ fits very well to the conjecture, although with certain oscillations on top of the straight line.  We believe these oscillations to be of the same origin as already seen in the non-interacting case, i.e. related to quenching on the counting field, however this still requires further investigation.  The real part of $\dot{F}$ also appears to converge to the line given by the conjecture after some initial transients.  There seems to be some evidence that the closer $V_\mathrm{SD}$ is to $V_c$, the longer one has to wait before convergence -- however this also requires further investigation.

We turn now to the next two plots for $V_\mathrm{SD}=0.37$ and $\mathrm{SD}=0.39$ which is just below and just above the critical bias $V_c$.  Just below $V_c$, one sees a continuation of the previous trend -- $\textrm{Im}(\dot{F})$ fits well to the conjecture though with additional oscillations; while $\textrm{Re}(\dot{F})$ looks like it will converge to the line of the conjecture, although in this case there is not enough data to say this for certain.  The plot for $V_\mathrm{SD}=0.39$, just greater than $V_c$ is however very revealing.  By comparison with the previous plot, we see that the analytic results have jumped, due to the discontinuity at $V_c$.  However, the numerical results are very close to the previous panel for $V_\mathrm{SD}<V_c$; in other words the evolution of the numerical (finite-time) results appears to be smooth as a function of $V_c$.  This gives strong evidence to support the idea that there is no discontinuity for any finite time -- as suggested in question one above.

Assuming that the infinite-time limit results are correct, we expect that the time evolution will eventually cross over to the analytic values according to the conjecture; the numerics suggest that this crossover will happen at a time scale 
\begin{equation}
t_m \propto (V_\mathrm{SD}-V_c)^{-1}.
\label{eq:critscale}
\end{equation}  
This statement is further supported by the final two panels of Fig.~\ref{fig:TimeSlices} which shows the time evolution at slightly larger times.  Here, we see that the short(er) time behaviour (which is similar to an analytic continuation of the CGF below at $V_\mathrm{SD}<V_c$) eventually evolves to reach strong peaks in $\dot{F}$ in both the real and imaginary parts.  The location of these peaks has already from Fig.~\ref{fig:coolplot} been seen to be at a time given by \eqref{eq:critscale} above.  We assume that at times (significantly) longer than the location of these peaks, the evolution eventually settles to the analytic long-time values -- certainly for larger values of $V_\mathrm{SD}$ the numerics and analytics are in good agreement (c.f. Fig.~\ref{fig:LargeVoltage}), although at these larger values the strong peaks are not seen in the time evolution.  Clearly more data is necessary to make any definitive conclusions; we defer this to a future work.

\begin{figure*}
\begin{center}
\includegraphics[width=0.49\textwidth,clip=true]{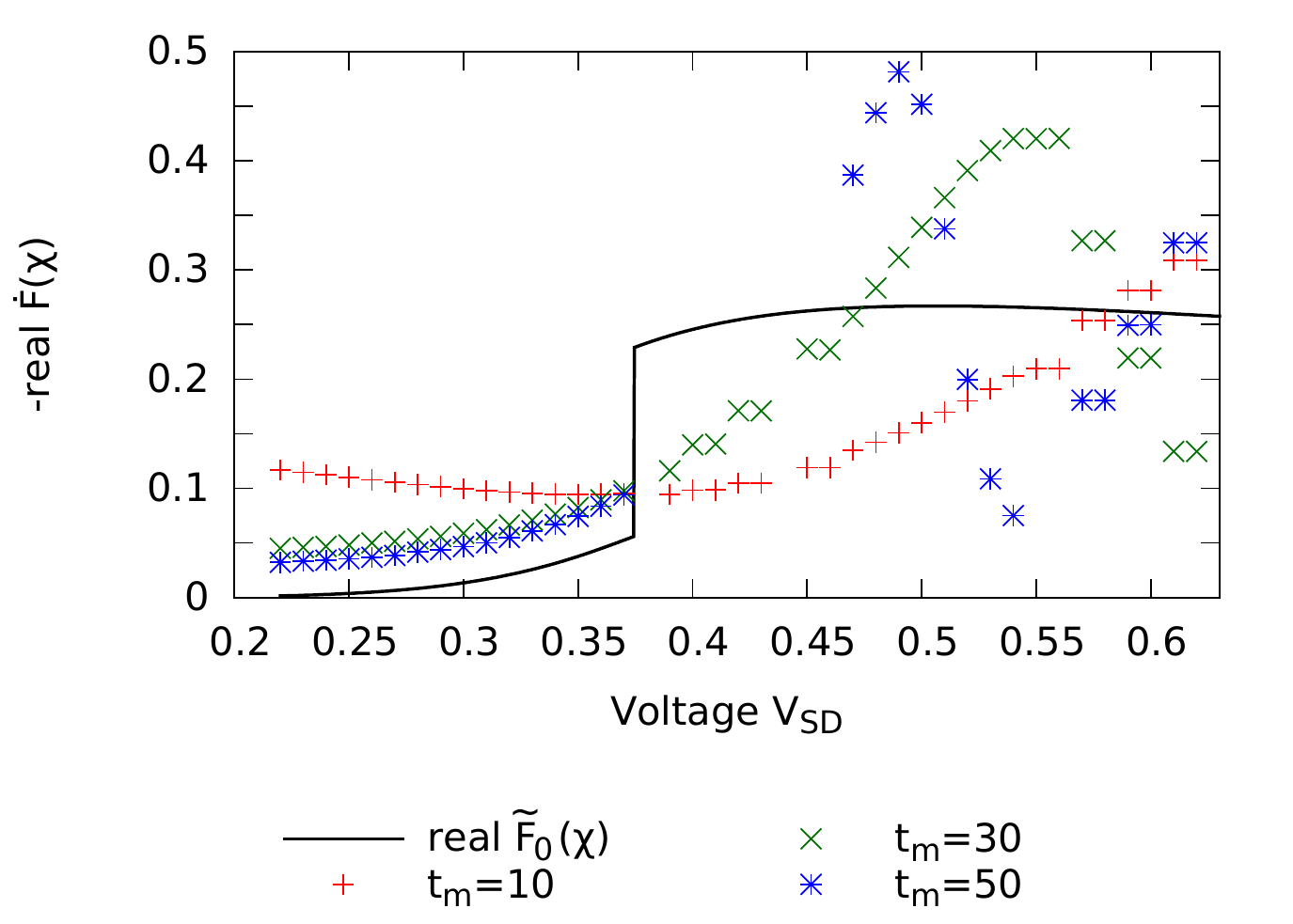}
\includegraphics[width=0.49\textwidth,clip=true]{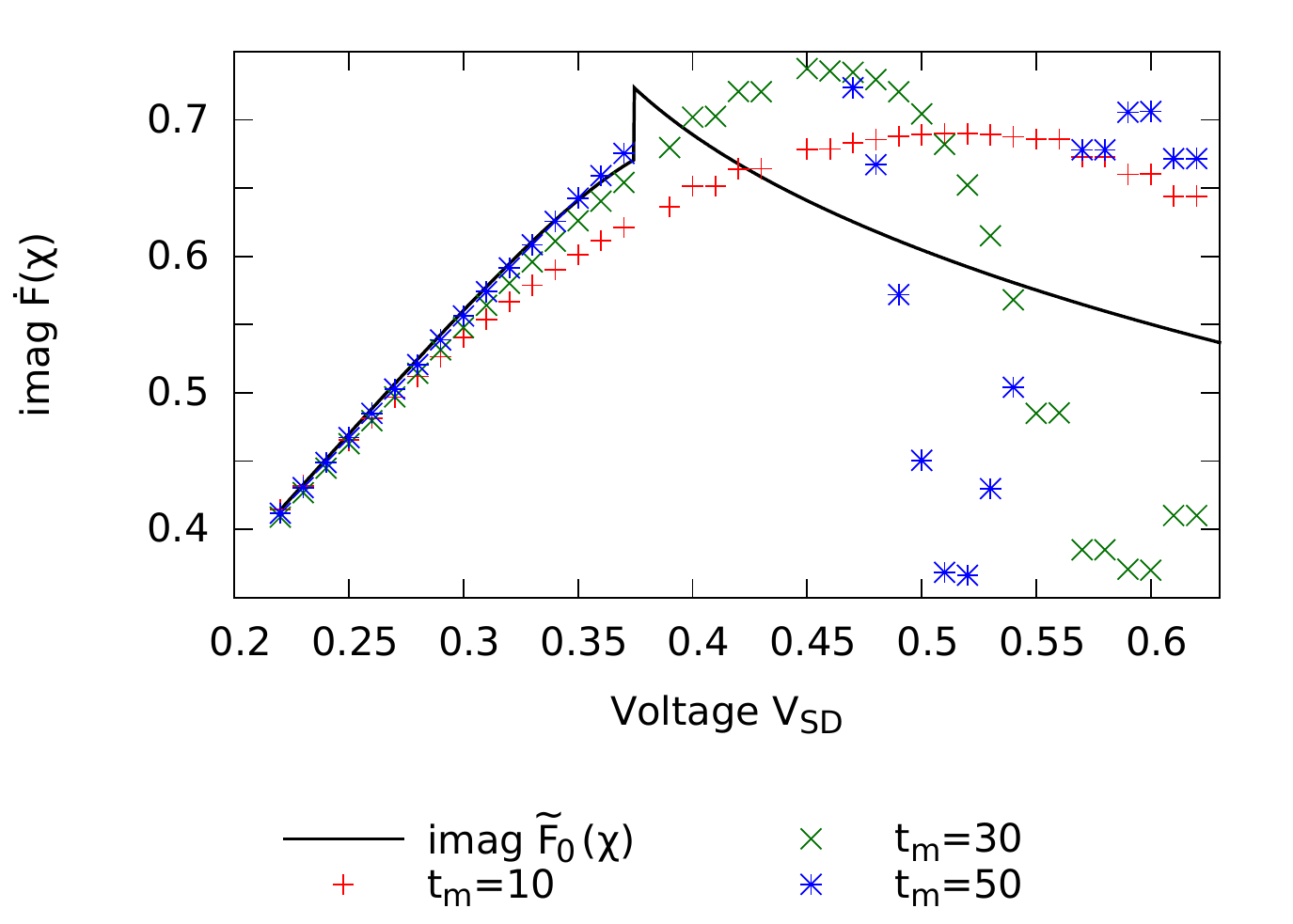}
\end{center}
\caption{The real (left) and imaginary (right) parts of $\dot{F}$ as a function of $V_\mathrm{SD}$ for $\chi=0.6\pi$ and various values of measuring time $t_m$.  The solid line is the analytic result $\tilde{F}_0$ which is the limit $t_m\rightarrow \infty$, and shows a discontinuity at $V_\mathrm{SD}=V_c\approx 0.374$.}
\label{fig:FvV}
\end{figure*}

As an alternative visualisation, we can take horizontal cross sections of Fig.~\ref{fig:coolplot}, in other words, plot $\dot{F}$ as a function of $V_\mathrm{SD}$ for given values of measuring time $t_m$.  This is shown in Fig.~ \ref{fig:FvV}.  The long time limit $\tilde{F}_0$ shows discontinuities at $V_\mathrm{SD}=V_c$, however the numerical results from different finite times appear to all be smooth functions of $V_\mathrm{SD}$.  For $V_\mathrm{SD}<V_c$, the time sequence converges towards the analytic long-time limit -- in fact, as shown in Fig.~\ref{fig:TimeSlices}, it does so according to the conjecture presented in this work.  For $V_\mathrm{SD}>V_c$ however, the finite time results develop additional structure in the form of peaks that become narrower as $t_m$ is increased.  One can certainly imagine how these peaks tend to the discontinuous analytic result as $t_m\rightarrow \infty$; however more data is necessary to investigate this further.

Finally, let us comment briefly on the role of finite-time on the measurement of fractionally charged quasiparticles.  The background of this problem is that if one does a projective measurement of the number of electrons in the lead, one must obtain one of the eigenvalues of the number operator, i.e. an integer.  The measurement of fractional charge is therefore intimately related to the quantum mechanical measurement problem.  We will not review this in any detail here, but would like to add an extra ingredient to the discussion: the role of finite measuring time.  In Section \ref{sec:bif}, we showed that, at least for the interacting RLM, the fractionalisation of charge occurs mathematically via a bifurcation at a branch point of the CGF.  We also demonstrated that at finite times, the CGF appears to be a smooth function.  At present, our data does not allow us to hypothesize on the periodicity of $\dot{F}$ as a function of measuring time -- however we believe this is an interesting question that could shed some light on the experimental detection of fractional charge.

\section{Summary}
\label{sec:end}

To summarise, we have discussed the cumulant generating function $F(\chi,t_m)$ of full counting statistics, and in particular the evolution of this function with measuring time.  We demonstrated that the leading corrections to the long time limit $F\sim \tilde{F}_0 t_m$ are logarithmic in nature at zero temperature \eqref{eq:series}, and have conjectured that the coefficient of this subleading term is universally related to the long time limit via Eq.~\eqref{eq:conjecture1}.  While we have presented strong numerical evidence for this conjecture for one specific model, its more general validity is an open question.  One promising direction in this regard is based on a Keldysh expansion of the CGF \cite{us-future}.

Another open question which is of great importance for extrapolating numerical results to longer times is about the nature of other finite-time corrections to the CGF, such as the oscillations seen in Fig.~\ref{fig:RLM-time-evolution}.  While we don't expect these to be universal, it is important to understand their form and origin in order to fit numerical data to obtain reliable results about the long-time limit of strongly correlated models.

Our numerical test of the conjecture were on the self-dual interacting resonant level model, \eqref{eq:IRLM}, where we further demonstrated that the long-time limit of the CGF, $\tilde{F}_0$, has branch cuts associated with fractionalize of charge at some critical voltage $V_c$.  We demonstrated that the CGF at finite times however shows no discontinuities; but does show distinctive features in the time evolution associated with this critical point; occurring at times satisfying Eq.~\eqref{eq:critscale}.  We furthermore suggested that this may have strong implications in the attempted experimental measurement of fractional charge.  Understanding the physics at this non-equilibrium critical point is an open question, and we believe one of the most fascinating future directions in the theoretical study of non-equilibrium systems.

We thank A.~Komnik, D.~Bagrets and D.~Gutman for insightful discussions.  HS' work was supported by the French Agence Nationale pour la Recherche (ANR Projet 2010 Blanc SIMI 4 : DIME) and the US Department of Energy (grant number DE-FG03-01ER45908).

\appendix

\section{Real time numerical simulations}
\label{sec:Size}

To study transport properties in the real time numerical method \cite{PS:PRB04,PS:Ann2010}, one initially finds the ground state of the system subject to a non-uniform potential of $\pm V_\mathrm{SD}/2$ on the left (right) lead.  This charge imbalance potential is then quenched off at time $t=0$, and the system is evolved numerically according to the Schr\"odinger equation
\begin{equation}
| \Psi(t) \rangle = e^{-i{\cal H} t} | \Psi(0) \rangle.
\end{equation}
By calculating the time evolution of the state, one can then evaluate the desired operator.  In particular, the protocol by which one can obtain the CGF of FCS by this method was presented in Ref.~\onlinecite{CBS-2011}.

For interacting systems, we numerically evolve the system using the density matrix renormalization group (DMRG) \cite{DMRG1,DMRG2}, while for the noninteracting systems, we use a much more efficient method based on single-particle evolution and slater determinants \cite{Schoenhammer-2007}.  However these particular numerical methods of time evolution have one thing in common: they all work on systems of a finite size, $M$.  This means that the leads are not considered infinite and instead consist of $M_a \approx M/2$ lattice sites.

There are a number of consequences of evolution on a system of finite size \cite{Schmitteckert-Carr-Saleur-2014}, the most crucial being that there is a transit time $t_T=v_c M$ ($v_c$ being the Fermi velocity in the leads) after which the charge imbalance has flowed across the system, bounces of the boundary and starts coming back thus destroying the steady-state current flow.  This therefore provides a hard limit on how long the system may be evolved for; the numerical equivalent of the battery going flat.  The typical sizes (and therefore maximum times) that may be simulated on a computer are limited by computer resources, although clearly the single-particle basis for the non-interacting case means that much larger systems may be used there.

While we refer to previous work \cite{PS:PRB04,PS:Ann2010,Schmitteckert-Carr-Saleur-2014} for a more detailed discussion of the manifestations of finite-size in non-equilibrium transport, we mention here one extremely useful result that is used in these proceedings.  In a (conformally invariant) system of finite size, one can capture the leading effects of times close to the transit time when the current is beginning to reverse direction by replacing the measuring time with a \textit{conformal time} \cite{Schmitteckert-Carr-Saleur-2014}.  In the simulations, the counting begins at a time $t_0$, which is chosen to be sufficiently long after the initial quench at $t=0$ so that the largest transients have died away and the system is in a (quasi-)steady state.  The conformal substitution therefore amounts to replacing the measuring time
\begin{equation}
t_m=t-t_0
\end{equation}
by the conformal time
\begin{equation}
   d(t_m,M) = \left( \sin\frac{ \pi t }{M/v_c} - \sin\frac{ \pi t_0 }{M/v_c} \right) \frac{ M \pi}{v_c}.
   \label{eq:ConformalTime}
\end{equation}
For short times $t_m \ll t_T$, we find $d(t_m) \approx t_m$.  The difference only becomes important at times near the transit time.

Equation \eqref{eq:ConformalTime} is based on the usual conformal transformation between the plane and a cylinder of finite extent in imaginary time \cite{CFT-book}, and analytically continuing to real time.  We note that while this procedure is not entirely justified in this setup \cite{Cardy-Calabrese-2006}, a comparison to numerics shows that it captures the leading effect of the back-reflection off the edges of the system remarkably well \cite{Schmitteckert-Carr-Saleur-2014} .

\section{Interacting Resonant Level Model away from the self-dual point}
\label{sec:otherU}

\begin{figure*}
\begin{center}
\includegraphics[width=0.49\textwidth,clip=true]{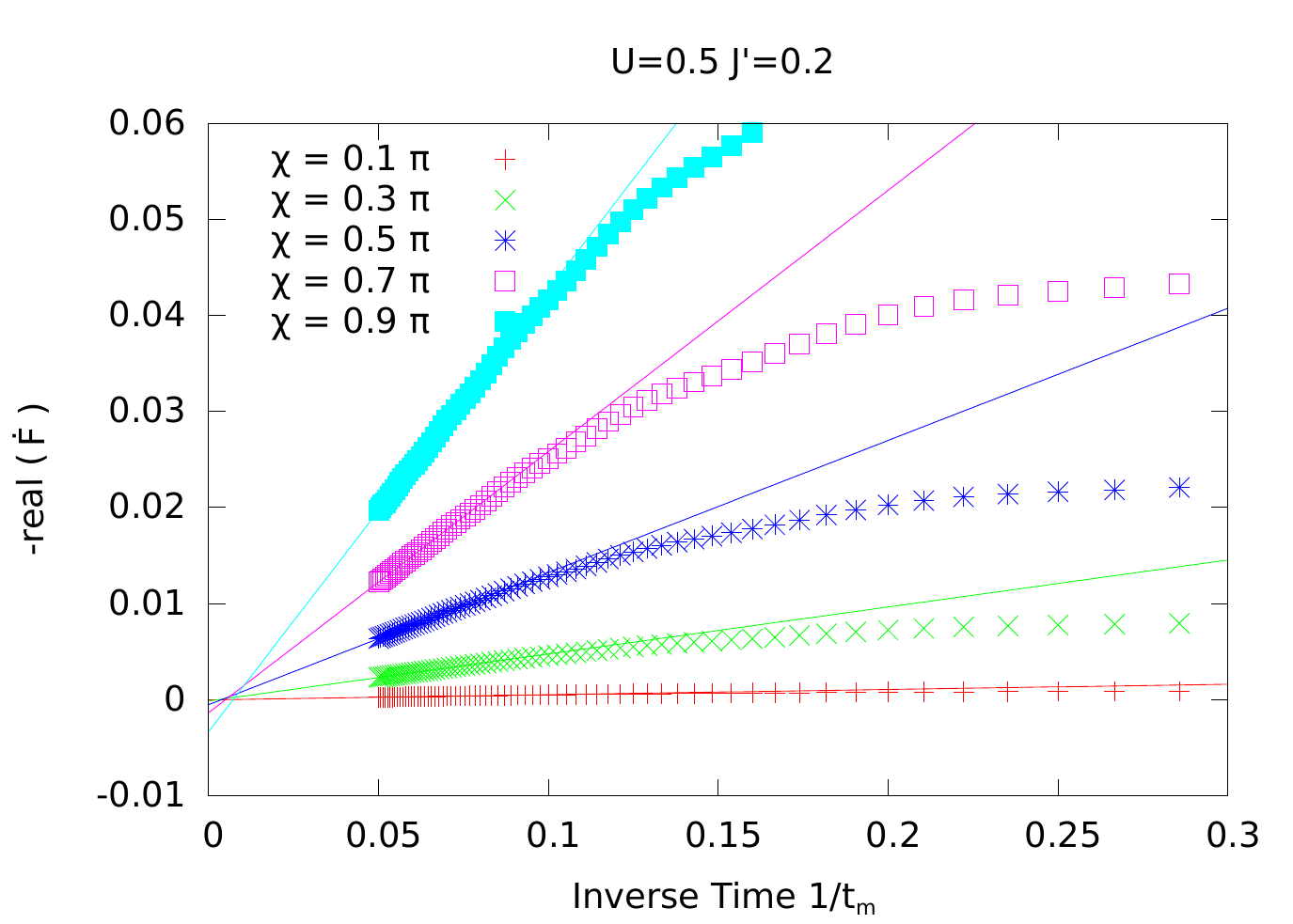}
\includegraphics[width=0.49\textwidth,clip=true]{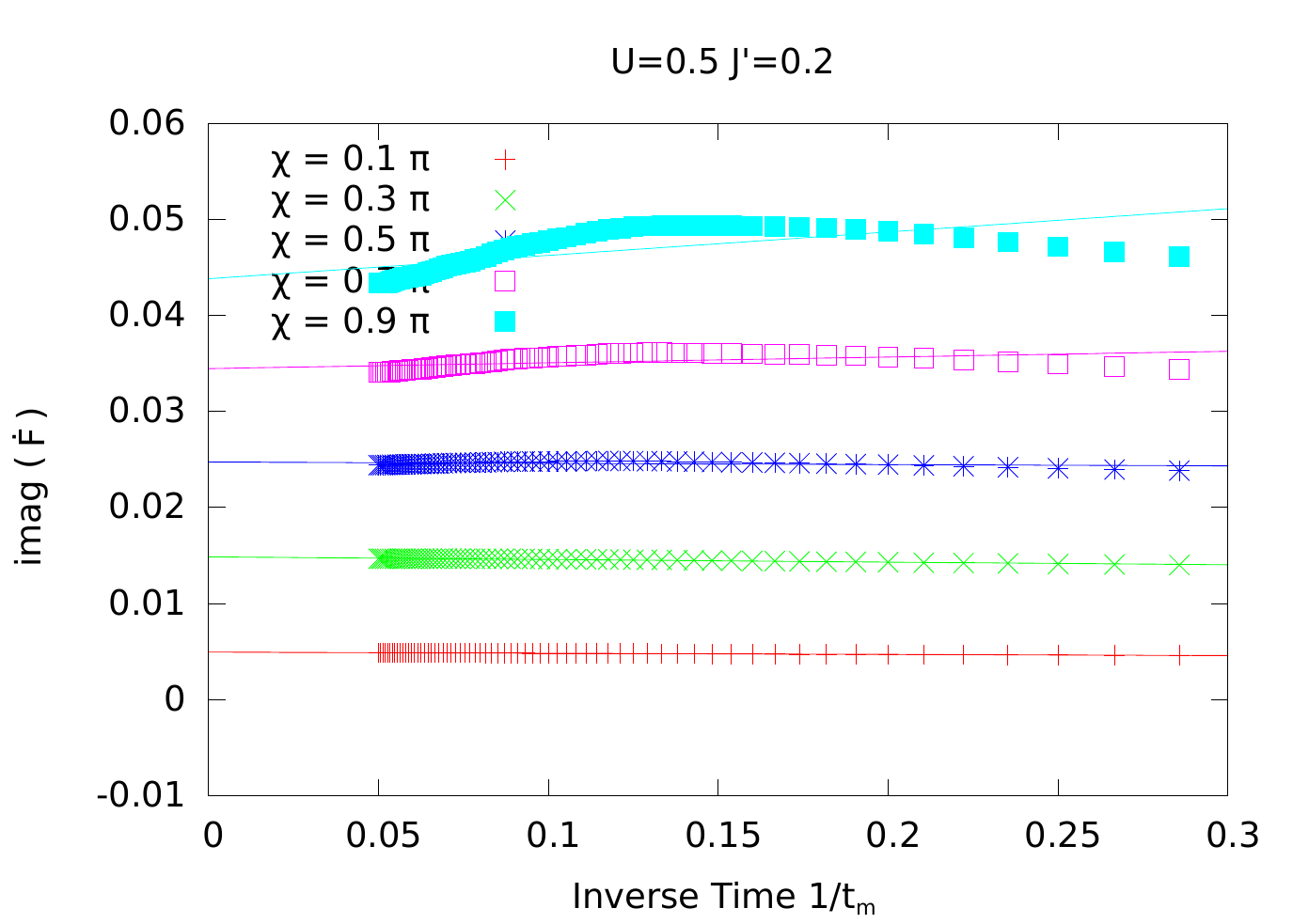}
\end{center}
\caption{Time evolution of the RLM \eqref{eq:IRLM} away from the self-dual point; this plot is for $U=0.5$.  The $1/t_m$ behavior can be clearly seen, thus demonstrating the validity of the series \eqref{eq:series}.}\label{fig:NonSelfDual}
\end{figure*}

While there is a lot of work on the FCS of other interacting models out of equilibrium (see e.g. \cite{Komnik-2007}), there are no exact results available for the CGF.  In particular, there is a lot of work on the interacting RLM, \eqref{eq:IRLM} for more general values of $U$ \cite{Andergassen-2011,Kennes-Meden-2013,Doyon-2007}, however without exact results for comparison to numerics, it is not easy to check the validity of the conjecture \ref{eq:conjecture2}.

This notwithstanding, the conjecture \ref{eq:conjecture2} relates the finite-time corrections $\tilde{F}_1$ to the long-time limit itself $\tilde{F}_0$ for any individual model.  Hence with enough data over a range of voltages, one can fit the numerical data to the series \eqref{eq:series}, numerically differentiate the obtained $\tilde{f}_0$ and hence check the relation \eqref{eq:conjecture2} completely numerically.  This requires a lot of high-quality numerical data and is work in progress.  Here we present only the first step of this, which is verifying that the corrections to the long time behavior of $\dot{F}$ are indeed proportional to $1/t_m$ (see Eq.~\ref{eq:series}) for the interacting RLM but \textit{away} from the self-dual point.

In Fig.~\ref{fig:NonSelfDual}, we show data for the measuring time evolution of the CGF for the interacting RLM at $U=0.5$.  This data shows clear $A+B/t_m$ behavior, thus demonstrating the validity of the series \eqref{eq:series}.  From the graphs of $\dot{F}$, it is also clear what is the numerical difficulty in testing the conjecture \eqref{eq:conjecture2}.  In particular, in the real part of $\dot{F}$, we see that we must extrapolate the numerical data from typical maximum times available over at least one order of magnitude to get the value of $\tilde{F}_0$.  As $\tilde{F}_0$ must be obtained with high precision in order to carry out a numerical derivative, this is a daunting task.  The results of this when complete will be presented elsewhere.


\end{document}